\documentclass{article}
\usepackage{graphicx} 
\usepackage[a4paper, total={6.47in, 10in}]{geometry}
\usepackage[utf8]{inputenc}
\usepackage[T1]{fontenc}
\usepackage[english]{babel}
\usepackage{subcaption,wrapfig}
\usepackage[dvipsnames]{xcolor}
\usepackage{amsmath}
\usepackage{amsfonts}
\usepackage{amssymb}
\usepackage{amsthm}
\usepackage{float}
\usepackage{braket}
\usepackage{physics}
\usepackage{braket}
\usepackage{bbold}
\usepackage{slashed}
\usepackage[font=small,labelfont=bf]{caption}
\usepackage{siunitx}
\usepackage{mhchem}
\usepackage{comment}
\usepackage{numprint}
\usepackage{lipsum}
\definecolor{ForestGreen}{rgb}{0.2,0.9,0.5}
\definecolor{mypink1}{rgb}{0.858, 0.188, 0.478}
\usepackage{url}
\usepackage{authblk}

\newcommand{\noise}{\text{noise-only scheme}}
\newcommand{\noisee}{\text{noise-only scheme }}
\newcommand{\nonlin}{\text{method}}
\newcommand{\nonline}{\text{method }}
\newcommand{\nonlins}{\text{methods}}
\newcommand{\nonlines}{\text{methods }}

\author[1]{Leen Mys \thanks{leen.mys@vub.be}}
\author[1]{Guy Verschaffelt}
\author[1]{Guy Van der Sande}

\affil[1]{Applied Physics Research Group, Vrije Universiteit Brussel, Pleinlaan 2, 1050 Brussels, Belgium}

\title{Predicting the optimal noise strength for solving optimization problems with analog Ising machines}

\begin{document}
	
	\maketitle
	
	\begin{abstract}
		Analog Ising machines are dedicated hardware solvers designed to solve NP hard optimization problems. 
		However, the global optimum is often not found as the system gets stuck in local minima. While several strategies exist to increase the chance of escaping local minima, often these methods needs extensive parameter tuning. In this work, we investigate the injection of large noise as a scheme on its own and in combination with annealing to improve the success rate and the time-to-solution (TTS) of analog Ising machines for MaxCut problems. We demonstrate that optimizing the noise improves the TTS by several orders and makes both approaches competitive with the state of the art, such as chaotic amplitude control. Moreover, we are able to predict a good noise value based on the problem connectivity and coupling strength, eliminating the need for costly parameter optimization. 
	\end{abstract}
	
	\section{Introduction}
	As the demand for energy-efficient and high-speed computational power continues to grow, interest increases in alternative computing systems to replace the traditional Von-Neumann system for specific high-demanding tasks. A set of computational tasks which in particular would benefit from faster and more efficient computing power are combinatorial optimization problems. A non-traditional approach to solve optimization problems in general is the so called Ising machine (IM), inspired by the NP-complete Ising problem. In an Ising problem the goal is to find the spin configuration of a magnetic material that minimizes its energy. An IM is a piece of hardware, where the Ising problem is recreated using artificial spins. It relies on the natural tendency of the Ising spins to evolve toward their ground state. So it naturally minimizes the Ising energy function
	\begin{equation}
		H_{\text{Ising}} = -\frac{1}{2} \sum_{ij}^{N} J_{ij} \sigma_{i} \sigma_{j} - \sum_i^{N} h_i \sigma_{i}
	\end{equation}
	where $J_{ij}$ is the coupling interaction between two spins, $N$ is the amount of spins in the problem, $h_i$ is the external field and $\sigma_{i} \in \{-1,1\}$ is the binary spin. The optimization problem to be solved is encoded in the coupling interaction $\boldsymbol{J}$ and the local field $\boldsymbol{h}$ while the ground state configuration is the solution of the problem. 
	It has been show that all combinatorial optimization problems can be mapped onto an Ising problem polynomially. There are already a scala of problems for which it is known how to map them efficiently on an Ising problem \cite{Lucas_Ising_2014}. For example, IMs can solve a broad range of real-life problems \cite{takesue_finding_2025} ranging from job scheduling \cite{rieffel_case_2015} to protein folding \cite{perdomo_construction_2008}, traffic flow optimization \cite{neukart_traffic_2017} and finance related problems \cite{orus_quantum_2019, johnson_quantum_2011}.
	
	Here, we concentrate on analog IM without local fields ($h_i = 0$). To facilitate optimization of an Ising problem, the spin amplitudes are allowed to take a real-value. We speak then about analog spin amplitudes indicated by $x_i$ in analog IMs. This analog spin system can be mapped back to the original binary spin system by taking the sign of the spin amplitudes ($\sigma_i = \text{sign}(x_i)$). When using analog spins, it needs to be checked that the ground state of the analog spin problem, to which the spins evolve naturally, is the same as the optimal energy of the original binary problem. Nonlinear terms are often employed to impose a binary structure on the analog spin during the dynamic evolution \cite{leleu_combinatorial_2017}. The choice of nonlinear term can also effect the efficiency of the search for the ground state \cite{bohm_order--magnitude_2021}.
	
	There exist a multitude of physical implementations of IM using electronic \cite{zhang_review_2024} or photonic \cite{mohseni_ising_2022} platforms. 
	Possible optical analog Ising implementations use (degenerate) optical parametric oscillators (DOPO) \cite{mcmahon_fully_2016,inagaki_large-scale_2016,inagaki_coherent_2016,haribara_computational_2016,takesue_finding_2025},  opto-electric oscillators (OEO) \cite{bohm_poor_2019,cen_large-scale_2022} or optical Kerr resonators \cite{quinn_coherent_2024}.
	Different physical implementations induce different nonlinear terms, thereby influencing the efficiency of the IM \cite{bohm_order--magnitude_2021}.
	A major challenge in analog IM is that the spins can become trapped in a local minimum, preventing convergence to the sought global optimum \cite{leleu_combinatorial_2017}. Several factors or techniques can limit the chance of getting trapped, including annealing \cite{finnila_quantum_1994,johnson_quantum_2011}, tweaking nonlinearities \cite{bohm_order--magnitude_2021}, stochastic perturbations through noise injections \cite{pierangeli_noise-enhanced_2020,shi_enhancing_2024}, simulated bifurcation \cite{kashimata_efficient_2024,goto_high-performance_2021} or using amplitude control schemes \cite{leleu_destabilization_2019}. 
	
	While both annealing and adding noise are commonly used in analog IM \cite{bohm_order--magnitude_2021, lamers_using_2024}, it is not clear when one is beneficial over the other. Furthermore, recent studies have shown that injecting high levels of noise can be beneficial in improving the performance of IM in general \cite{liao_quantum_2022,liao_overdamped_2024, roques-carmes_heuristic_2020,lee_noise-augmented_2025,lee_correlation_2025}, and analog IM in particular \cite{pierangeli_noise-enhanced_2020, shi_enhancing_2024}. However, in most of these studies, noise is treated only as a tunable parameter for optimization, without deeper investigation into why it optimizes the time-to-solution and success rate or how the effect of noise depends on other aspects of the system. 
	The presence of an optimal noise level has been reported in \cite{shi_enhancing_2024}, but was not further explored and was only demonstrated for the success rate. 
	We would like to predict the optimal noise level from basic system characteristics, such as graph size or coupling strength, rather than relying on exhaustive parameter scans or Bayesian optimization methods. While these methods can be effective, they are either computationally expensive or require prior knowledge of the ground-state energy, which may not be available in practice. 
	
	In this work, we will investigate the role of noise injection in analog IM, both  as a scheme on its own and in combination with annealing. We benchmark the performances of the models including noise against the considered state-of-the-art chaotic amplitude control (CAC) IM using MaxCut problems to better understand the effect of noise in both situations. Additionally, we explore how the optimal noise regime depends on problem characteristics, particularly the graph size and the average number of connections per spin. We find a relationship between the noise and some basic system parameters, allowing us to predict the required optimal noise. 
	
	In section \ref{modenbench}, we will introduce the different schemes, \nonlines and benchmarks that have been used in our investigation. We then use one benchmark as an example to illustrate the influence of noise and investigate the relationship between different system parameters in section \ref{influencenoise}. Section \ref{comparison} will dive deeper in the performance differences between the different models. After that, we will investigate the relationship between noise and the system parameters in section \ref{relatie}. We will do this separately for small problems ($N \lesssim 100$) and large problems ($N>100$). Afterwards, we will verify the found relationship between noise and the system parameters and conclude the paper with some discussion and future work in the section \ref{discussion}. 
	
	\section{Models and benchmarks} \label{modenbench}
	An analog IM is a (physical) system of interconnected analog spins that has the tendency to evolve naturally towards the configuration of minimal energy. This evolution can be modeled by the differential equation
	\begin{equation}
		\frac{d x_i}{d t} = G \Big[x_i, \alpha, \beta \sum_{j=1}^{N} J_{i,j} x_j,  \gamma F(t) \Big]
	\end{equation}
	Here, $G$ is the nonlinear transfer function, $\alpha$ is the linear gain,  $\beta$ is the coupling strength, $x_i \in \mathbb{R}$ is the analog spin amplitude, $t$ is the re-normalized time, and $\gamma F(t)$ is a Gaussian white noise term with $\left\langle F(t) \right\rangle = 0$ and $ \langle F(t) F(t') \rangle = 2 \delta (t-t')$.
	As is described in \cite{bohm_order--magnitude_2021}, the transfer functions depends on the specific implementation and hardware of the IM. As this paper is only based on simulations and theory, we use  transfer functions which minimize the number of parameters, as to ease the analysis. Therefore, we will use the so-called polynomial and clipped \nonlin, where it is possible to eliminate $\alpha$ by (re-)defining $x_i$, $t$ and $\beta$, as described in \cite{lamers_using_2024}. The differential equation for the polynomial \nonline is 
	\begin{equation} \label{polynomial}
		\frac{dx_i}{dt} = -x_i - x_i^3 + \beta \sum_{j=1}^{N} J_{i,j} x_j + \gamma F(t) \\
	\end{equation}
	This is the more basic \nonlin, as it is the normal form of the pitchfork bifurcation of the spins. The polynomial \nonline also describes the traditional optical setup of a coherent IM using DOPO \cite{mcmahon_fully_2016,inagaki_large-scale_2016,inagaki_coherent_2016,haribara_computational_2016,takesue_finding_2025}. Futhermore, it is the Taylor expansion of the other \nonlins, such as the periodic \nonline \cite{bohm_order--magnitude_2021}. Here, it will be used to provide us with the necessary insight in the working of noise. The performance of the clipped \nonline was substantially better than the performance of the polynomial \nonlin. Its performance was comparable with the other \nonlines investigated in Ref. \cite{bohm_order--magnitude_2021}. 
	The differential equation for the clipped model is
	\begin{equation} \label{clipped}
		\frac{dx_i}{dt} = 
		\begin{cases}
			-x_i - x_i^3 + \beta \sum_{j=1}^{N} J_{i,j} x_j + \gamma F(t) & x_i \leq x_{cutoff} \\
			 0 & x_i > x_{cutoff} \\ 
		\end{cases}
	\end{equation}
	We have investigated several values for $x_{cutoff}$ and its effect on the performance of the IM. The exact value does not seem to be important,as long as $x_{cutoff} < 2$ (see supplementary materials section \ref{clip}). The $-x_i^3$ term is sometimes omitted in the literature for the clipped \nonlin, but adding or omitting this term seems not to influence the performance. Here we choose to keep this term and take $x_{cutoff} = 1$. 
	
\begin{figure}
		\centering
		
		\includegraphics[width=0.8\linewidth]{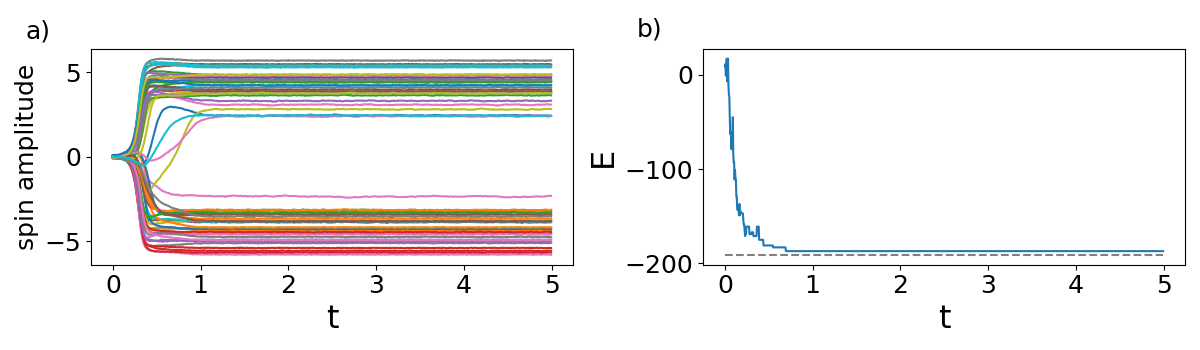}
		\includegraphics[width=0.8\linewidth]{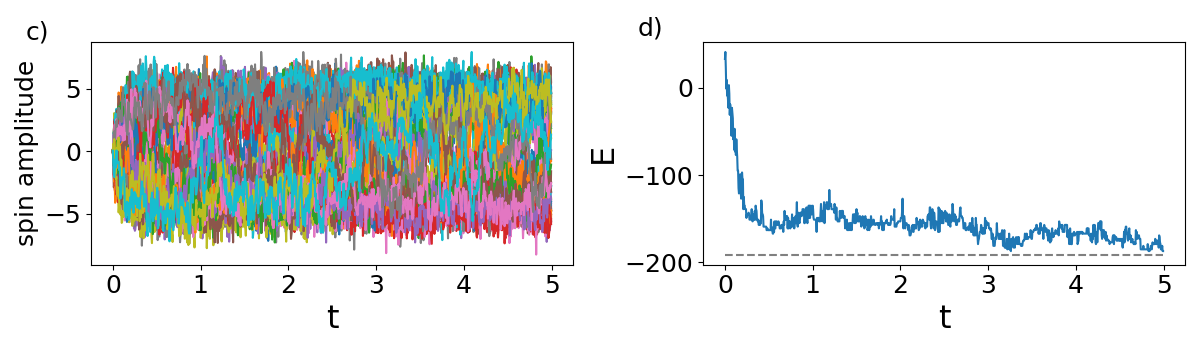}
		\caption{(a,c) The spin amplitudes and (b,d) energy evolution of g05\_60.3 for $\beta = 3$ and $\gamma = 0.1$ or $\gamma = 9.4$ respectively. Both trials are run for $t =75$ and $dt = 0.005$.  The ground state (grey dotted line; $E = -191$) is reached at $t = 4.91$ for $\gamma = 9.4$ and never for low noise. }
		\label{Evolution}
	\end{figure}

	Eqs.~\eqref{polynomial} and \eqref{clipped} are solved numerically, to simulate the natural evolution of the spin system towards the ground state. This evolution is shown in Fig.~\ref{Evolution} for problem g05\_60.3 of the BiqMac library. In Fig.~\ref{Evolution}(a) the time evolution of all the 60 spin amplitudes is demonstrated over time for $\beta = 3$. Starting from their almost zero position, the spin amplitudes will evolve naturally to a clear 'up' or 'down' position.  Notice that there is a substantial difference between the amplitudes of the spins, often refered to as amplitude inhomogeneity. Amplitude inhomogeneity increases the chance of the spins getting stuck in non-optimal spin configurations \cite{leleu_combinatorial_2017}. 
	The corresponding evolution of the energy of the system is shown in Fig.~\ref{Evolution}(b). The energy level quickly drops to a level close to the  ground state energy, here shown by the grey dotted line. In this example, the system never reaches the actual ground state energy, but is trapped in a local minimum just above it.  In Fig.~\ref{Evolution}(c-d), the spin amplitudes and energy evolution is shown for a system with more noise. While $\gamma = 0.1$ for Fig.~\ref{Evolution}(a-b), it is increased to $\gamma = 9.4$ for (c-d) to demonstrate the effect of large noise, which means that $\gamma \gtrsim \beta$. Now there is no clear distinction between up and down spins in the spin amplitudes evolution. While initially the energy also drops quickly, it keeps fluctuating, finally reaching the ground state at $t = 4.91$.
	
	We will evaluate the performance of an IM with a set of fixed parameters, by simulating several runs, each starting from random initial conditions. We will rely on two metrics to quantify the performance: the success rate (SR) (number of runs in which the ground state or target state was reached) and the time-to-solution (TTS). The time-to-solution is calculated as following
	\begin{equation}
		TTS = \frac{1}{N_{\text{success}}} \sum_i t_i \frac{\log{0.01}}{\log{1-\text{SR}}}
	\end{equation}
	where $N_{\text{success}}$ is the number of runs in which the ground state energy was reached and $t_i$ the simulation time it took in the i$^{th}$ successful run to reach the ground state energy. For practical reasons, our simulations have a runtime limit, which consists of 15000 numerical iterations for small problems ($N \lesssim 100$) and 20000 iterations for large problems ($N>100$). Such one iteration consists of updating all the spins amplitudes once using the spin amplitudes of previous iteration. For the first iteration, the spin amplitudes are randomly initialized using an uniform distribution with as boundary $|x_i| < 0.1$. This is also visible in Fig.~\ref{Evolution}(a,c).
	
	To optimize the performance of an IM, several schemes are proposed in the literature. Most of these schemes are focused on avoiding to get trapped in local minima, as this problem drastically increases the TTS and lowers success rate \cite{bohm_order--magnitude_2021, dobrynin_energy_2024}. Here, we will focus on the following three schemes: 
	The first scheme relies on the injection of noise in the system to perturb the spin amplitudes to escape local minima. The noise can induce spin flips even when this is not energetically favorable, allowing the system to explore a broader region of the phase space. This scheme will be referred to as \noisee.
	
	In the second scheme, $\beta$ will be annealed linearly, i.e. $\beta = \beta_0 + v *t$ where $v$ is the annealing speed. The working of this scheme was partly explored in \cite{lamers_using_2024}, where it was shown that for some problems annealing always leads to the desired ground state as long as the annealing speed is kept low enough. These problems were therefore labeled “Ising easy” in \cite{lamers_using_2024}. There are, however, also problems for which annealing leads to very low SRs, certainly so if the noise strength $\gamma$ is low. These problems are called “Ising hard” in \cite{lamers_using_2024}, and noise is crucial to effectively solve these problems. Notice that if a problem is Ising easy or hard, depends on the chosen \nonline and parameters. For example, it is possible that a problem is Ising hard for the polynomial \nonline and Ising easy for the clipped \nonlin.
	
	The third and last scheme we include in our analysis is chaotic amplitude control (CAC) which was introduced by T. Leleu, \textit{et al.} in \cite{leleu_destabilization_2019}. CAC will be used as baseline for the other two schemes, as it is one of the most highly-performant analog IM. The parameters needed for this scheme will be tuned using Bayesian optimization \cite{martinez-cantin_bayesopt_2014} for $\beta = 1$. Afterwards, a scan over $\beta$ will be made similar to the scan done for the two other schemes, using the optimized parameters for $\beta = 1$. As this scheme is inherently chaotic, it is not beneficial to add noise, according to \cite{leleu_scaling_2021}. 
	
	For the two first schemes, the influence of noise will be investigated, aiming to predict the best noise value for a certain problem. This will be done using simulations of MaxCut problems, where the cut number $\frac{1}{4} \left( \sum_{ij} J_{ij} - \sum_{ij}^N J_{ij} \sigma_i \sigma_j \right)$ needs to be maximized when separating a graph structure in two parts. This is related to minimizing the Ising energy and therefore this type of problems can be easily solved by IM. Maximizing the cut number and minimizing the energy will be used interchangeable in the remainder of the paper. We use instances from Biq Mac library \cite{noauthor_biq_nodate} and the SuiteSparse Matrix Collection \cite{noauthor_gset_nodate}. Furthermore, extra problems are generated using rudy \cite{noauthor_index_nodate}, a random graph generator that was also used to generate the MaxCut instances from the BiqMac library and Gset.
	
	

	\section{Influence of noise} \label{influencenoise}
	Here, we study the effect of Gaussian noise on both the polynomial and clipped \nonlins, with the objective of maximizing the success rate and minimizing the time-to-solution (TTS). When these two goals do not align,  the TTS is prioritized, as we consider it to be the most relevant metric for assessing the efficiency of IM. 
	
	\begin{figure}
		\includegraphics[width=0.99\linewidth]{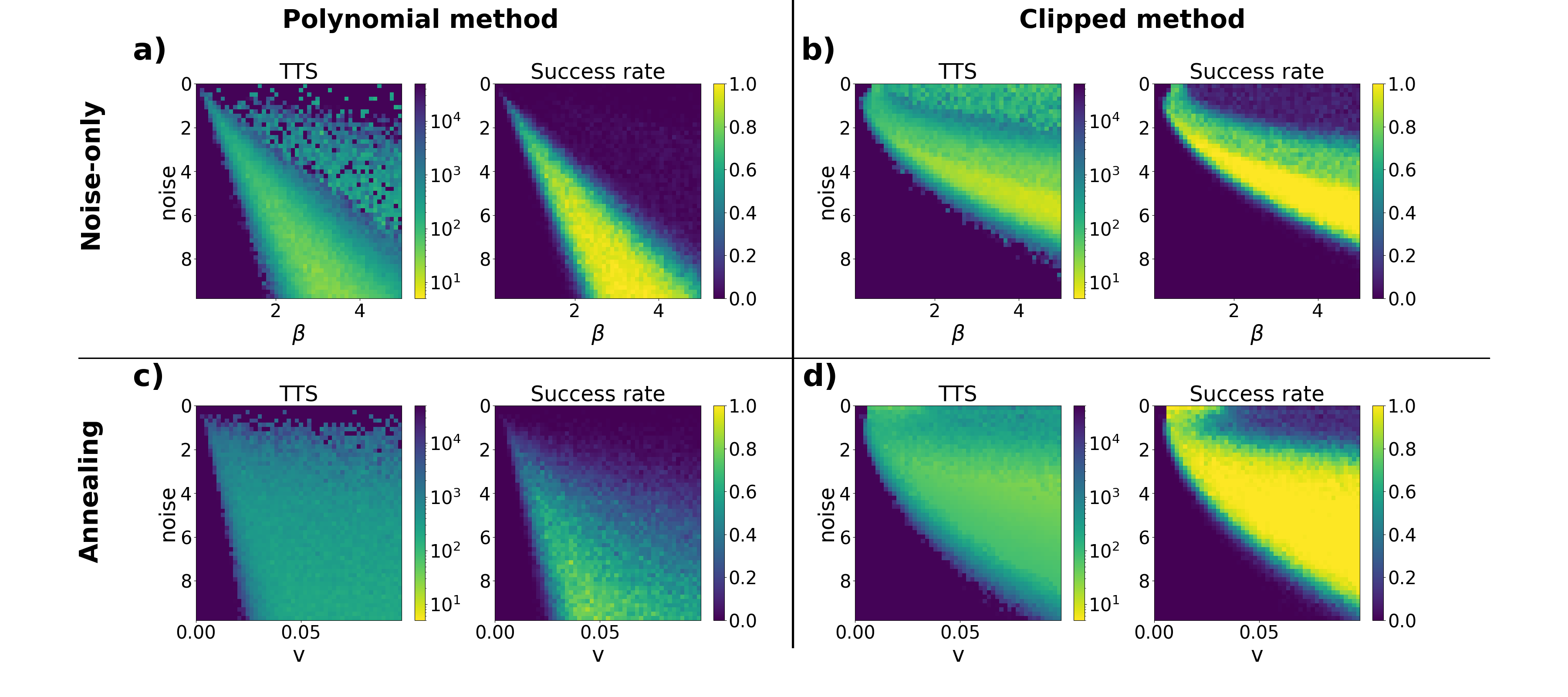}
		\caption{The success rate and the time-to-solution in function of $\gamma$  and $\beta$ for the \noisee (a,b) and $\gamma$ and $v$ for annealing (c,d) as heat map. The results for the polynomial \nonline can be seen at the left (a,c) and the results for the clipped \nonline at the right (b,d).}
		\label{g05_60.3}
	\end{figure}
	
	To illustrate the impact of the noise strength, we begin by examining the MaxCut instance g05\_60.3 of the Biq Mac library \cite{noauthor_biq_nodate}. This problem is selected as it clearly shows  that there is a relationship between noise and coupling strength, is not Ising easy for the polynomial \nonline \cite{lamers_using_2024} and is representative for the other problems studied. 	
	In Fig.~\ref{g05_60.3}, we present parameter sweeps over $\beta$, $\gamma$, and $v$ for several models as heat maps. The TTS and success rate for a scan over $\beta$ and $\gamma$ using the \noisee and polynomial \nonline is shown in Fig.~\ref{g05_60.3}(a). The lowest TTS's are found for large $\gamma$ and large $\beta$, which we define here as $\beta >1$. For example, $\beta = 3$ and $\gamma \approx 9.4$ results in a low TTS and a success rate of 0.99 for the polynomial \nonlin, which is also demonstrated in Fig.~\ref{Evolution}(b). Additionally, we see that for each $\beta$, there is an optimal noise that maximizes the success rate and minimizes the TTS. So, when keeping $\beta$ fixed and changing $\gamma$, the increase in success rate and decrease in TTS is more drastically seen for higher values of $\beta$, going from a success rate of zero to (almost) 1. There even seems to be a linear relationship between $\beta$ and the optimal noise for that coupling strength. When following this linear relationship from low $\beta$ to high $\beta$, the success rate and TTS are further optimized, which agrees with our first observation that the lowest TTS's are found for high $\gamma$ and $\beta$. This linear relationship will be investigated in more detail later in section \ref{relatie}.  
	
	In Fig.~\ref{g05_60.3}(b), the heatmap shows a scan over $\beta$ and $\gamma$ for the TTS and success rate for the clipped \nonlin. Again there is an optimal noise per $\beta$ and the TTS decreases for high $\beta$ and $\gamma$. However, the linear relationship between $\beta$ and $\gamma$ does not exist anymore. Here, it seems to be replaced by a square root behaviour, which we will verify later in section \ref{relatie}. Due to this different relationship, the noise values which lead to low TTS are not as high as for the polynomial \nonlin. 
	
	Similar results are found for annealing in  Fig.~\ref{g05_60.3}(c) for the polynomial \nonline and in  Fig.~\ref{g05_60.3}(d) for the clipped \nonlin. Here, the parameter scan is over $\gamma$ and $v$, as $\beta$ increases over time during a run. A comparable region is explored, as $\beta_0 = 0$ and $\beta$ goes up to 7.5 within the time limit for the annealing scheme for the highest speed ($v = 0.1$). There is a linear relationship for the polynomial \nonline and a square root relationship for the clipped \nonline similar to the \noise, but this time in function of $v$. 
	
	The existence of an optimal noise level is to be expected as follows. With insufficient noise, spin flips are rare, preventing effective exploration of the phase space and causing the system to remain trapped in local minima. Conversely, excessive noise drives the system into a Boltzmann sampling regime \cite{bohm_noise-injected_2022}, reducing the chance to reach the ground state. Therefore, an intermediate, optimal noise level exists, depending on $\beta$.
	A similar effect is seen when annealing. If the velocity is too low in comparison to the noise, the time limit of our simulations will prevent $\beta$ to reach high enough values to escape the region of Boltzmann sampling. However, if the speed is too high, $\beta$ will become too fast too high for the used noise to let the IM explore the phase space.
	
	In this example shown in Fig.~\ref{g05_60.3}, the TTS is significantly lower for the \noisee than for annealing with noise.
	While the MaxCut problem g05\_60.3 is used to showcase the importance of noise in the TTS and success rate, almost identical behaviour is observed in all the investigated MaxCut problems (see section \ref{extraprob}  in supplementary materials for other examples). Although not all problems achieve near-perfect success rates, nearly all exhibit an optimal noise level for each $\beta$, except for the most difficult problems which achieve (almost) no successful runs.
	
	\section{Comparison of different schemes and \nonlins} \label{comparison}
	Several \nonlines exist to prevent the analog IM to get stuck in local minima when solving Ising problems. In this section, we will compare annealing and \noisee to CAC, a highly-performant \nonline that we use as baseline \cite{leleu_scaling_2021}. From this point forward, we focus on TTS. Even when multiple models achieve a near-unity success rate, their TTS can differ substantially.
	
	\begin{figure}
		\centering
		\includegraphics[width=\linewidth]{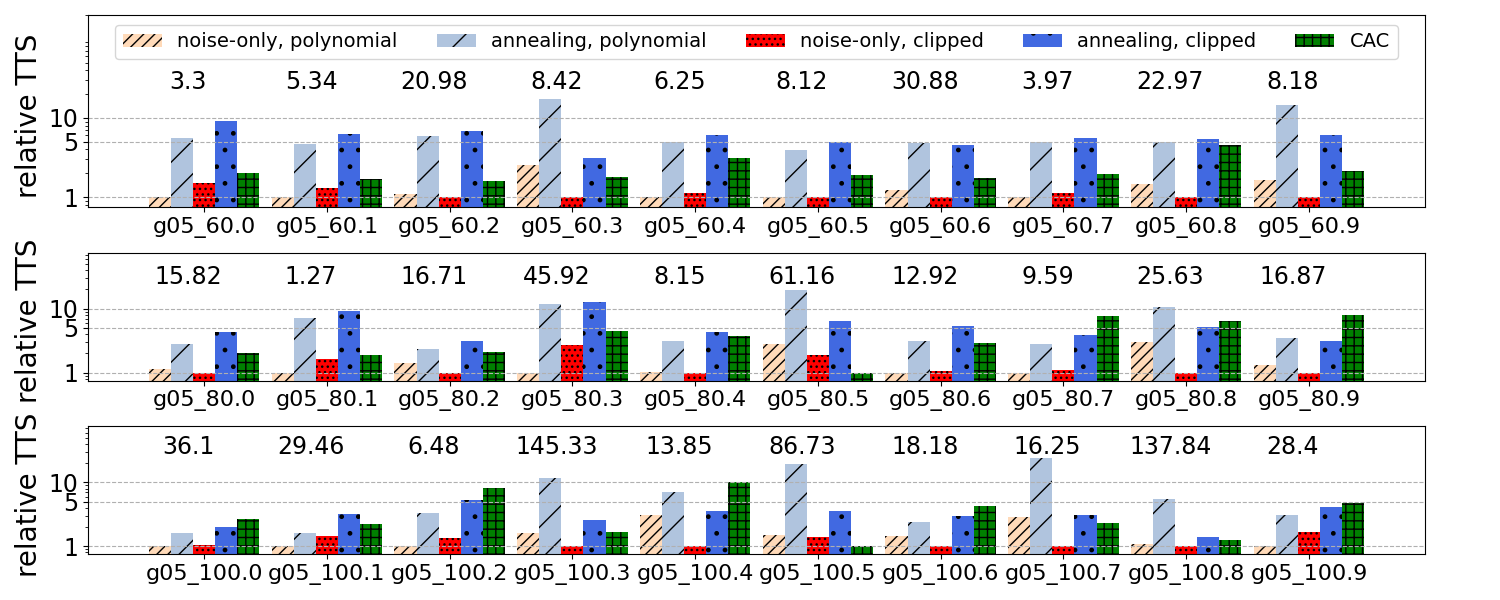}
		\caption{Time-to-solution of the \noisee and annealing scheme for the polynomial and clipped \nonline and CAC relative to the best TTS of these models. The TTS of the best model is mentioned above the bars to give an idea of the timescale.  All the problems are from the BiqMac library. }
		\label{TTS_dif}
	\end{figure}
	
	We compare the TTS for the polynomial \nonline and the clipped \nonline for all the g05 Biq Mac problems \cite{noauthor_biq_nodate} and G1-40 problems of the GSet \cite{noauthor_gset_nodate}. For this, we use the best TTS per model of each problem obtained from a parameter scan similar to the one shown in Fig.~\ref{g05_60.3}.
	The differences between the best TTS found for all the models and the best TTS per model are shown for each g05 problem in Fig.~\ref{TTS_dif} on logaritmic scale. The annealing results are in blue, the results for the \noisee in red and the result for the Bayesian optimized CAC in green with a checkered pattern. We compare the TTS for 5 models: \noisee and annealing scheme with the polynomial \nonline and the clipped \nonline and CAC. To provide a sense of absolute performance, the best TTS value for each problem is displayed above the bars in Fig.~\ref{TTS_dif}. The \noisee always outperforms annealing for the polynomial as well as the clipped \nonlin. The best TTS is provided 16 times by a clipped \nonlin, 12 times by the polynomial \nonline and 2 times by CAC. So CAC rarely results in a better TTS for small problems ($N \lesssim 100$) such as the BiqMac problems.

	\begin{figure}
		\centering
		\includegraphics[width=\linewidth]{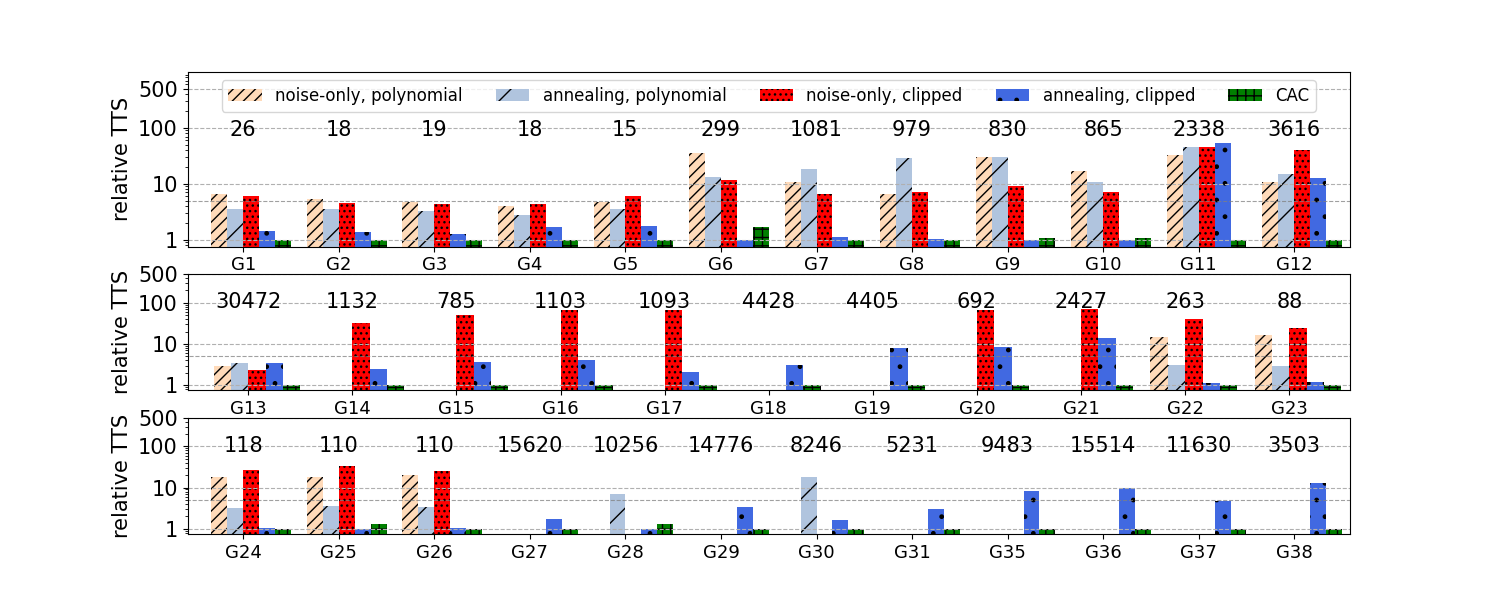}
		\caption{TTS of the polynomial and clipped \nonline for the \noisee and annealing scheme and of CAC relative to the best TTS of these models. The TTS of the best model is mentioned above the bars. When no bar is visible, 99.5\% of the best-known cut value is never reached by that \nonlin. All the problems are from the Gset. }
		\label{Gset_TTS}
	\end{figure}
	\begin{figure}
		\centering
		\includegraphics[width=\linewidth]{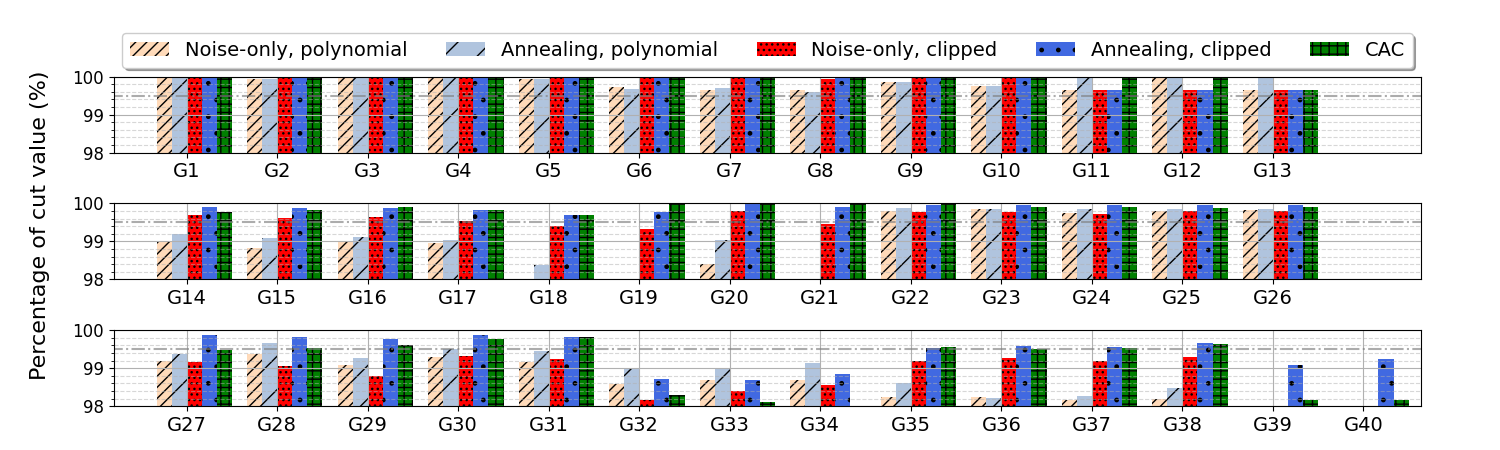}
		\caption{The highest percentage of the best-known cut value mentioned in the literature \cite{noauthor_gset_nodate} that is reached with annealing, the \noisee and CAC for the Gset. When the bar reaches 100\%, the highest found value and the best-known value are equal. }
		\label{Gset_diff}
	\end{figure}
		
	The relative best TTS of the \noisee and annealing scheme for the polynomial and clipped \nonline and of CAC for the Gset are shown in Fig.~\ref{Gset_TTS}. The different models are displayed with the same colors and patterns as in Fig.~\ref{TTS_dif}. Due to the difficulty level of these problem and the absence of a known ground state, the evolution is called a success if it reaches 99.5\% of the highest cut value mentioned in literature \cite{matsuda_benchmarking_2019}. There are 5 problems omitted (G32, G33, G34, G39,G40), as neither CAC, nor any of the other models reaches 99.5\% of the best cut value in the given simulation time. In Fig.~\ref{Gset_TTS}, we see that the best TTS is often given by CAC: it outperformed the other models in 28 of the 35 problems. However, the other models and specifically annealing with the clipped \nonline performs very similar to CAC: only in 8 of 35 cases is the TTS of the second best \nonline 5 times larger than the best TTS for CAC. When comparing  the \noisee with the annealing scheme, we see that annealing outperforms the \noisee in 32 of the 35 problems. Specifically annealing with the clipped \nonline gives the best results, which are often in the same order as CAC.  
	
	As the TTS is based on 99.5\% of the best-known cut value, it is also interesting to look at the highest cut value found within the time limit of $t =75$ during the parameter scan or Bayesian optimization. In Fig.~\ref{Gset_diff}, it is displayed which percentage of the best-known cut value is found for the \noise, annealing and CAC. It is the highest value found during the whole parameter scan or during the whole process of Bayesian optimization, so the parameters for which this value is found does not have to be the same as the parameters for the lowest TTS. When a bar reaches 100\%, it means that this model reached the best-known solution. CAC often reaches the best-known solution and reaches in general higher cut values than the other models. Annealing with the clipped \nonline performs almost equally good as CAC. Annealing mostly outperforms the \noise: annealing finds a lower energy than the \noisee in 37 out of the 40 problems. For 33 problems, this energy was found by the clipped \nonline.
	
	By comparing the results for the smaller BiqMac problems and the Gset, we get some contradictory results: for the BiqMac problems, \noisee always results in better TTS, but annealing reaches higher cut values and lower TTS's for the Gset. This indicates that annealing with large noise is a better scheme for larger and more difficult problem sets, such as Gset, whereas the \noisee performs better for simple tasks. The comparison with CAC shows that injecting large noise seems to speed up the process of finding the target. For small problems or when it is not important to find the lowest possible energy value, injecting noise is faster than CAC. For larger problems, annealing with the clipped \nonline obtains similar efficiencies as CAC. So, large noise significantly improves the performance of previously slower \nonlin, making them competitive with fast and promising approaches such as CAC.
	
	\section{Relation between the coupling strength / the annealing speed and the noise strength} \label{relatie}
	In the previous section, we have compared several \nonlines and schemes, using the minimum TTS obtained from parameter scans. We demonstrated that with optimal noise, these \nonlines are competitive with CAC. Now we will investigate which parameter values lead to the best TTS. We are particularly interested in the relation between the noise strength $\gamma$ and the coupling strength $\beta$. From Fig.~\ref{g05_60.3}, we noticed a roughly linear relationship between the noise and coupling strength for the polynomial \nonline and a square root dependence for the clipped \nonlin. From here on, we will focus on the \noise, as the relationship between $\gamma$ and $\beta$ looks more clear for the \noisee than for annealing in all the observed problems, as can be seen in Fig.~\ref{g05_60.3} and in Fig.~\ref{g05_N.0} from supplementary materials. This is because the area of low TTS is smaller and more delineated in the \noise. Additionally, the coupling strength $\beta$ and the annealing speed $v$ are related, so it will be possible to derive the relationship between the noise strength $\gamma$ and the annealing speed $v$ from the relationship between the coupling strength $\beta$ and the noise strength $\gamma$. It is therefore sufficient to investigate the relationship between $\gamma$ and $\beta$.	
	We will look at small ($N \lesssim 100$) and large problems ($N>100$) separately, as they have a different definition of success. For large problems, the best-known solution is never reached or only a few times, so it is not possible to analyze it. Therefore, the success rate is based on a target energy, which is a bit higher than the actual ground state energy. For the small problems, this is no issue. This difference in definition in success rate will influence the relation between the coupling strength and the noise strength.

	\subsection{Small problems} \label{relatiesmall}
	
\begin{figure}
		\centering
		\includegraphics[width=0.89\linewidth]{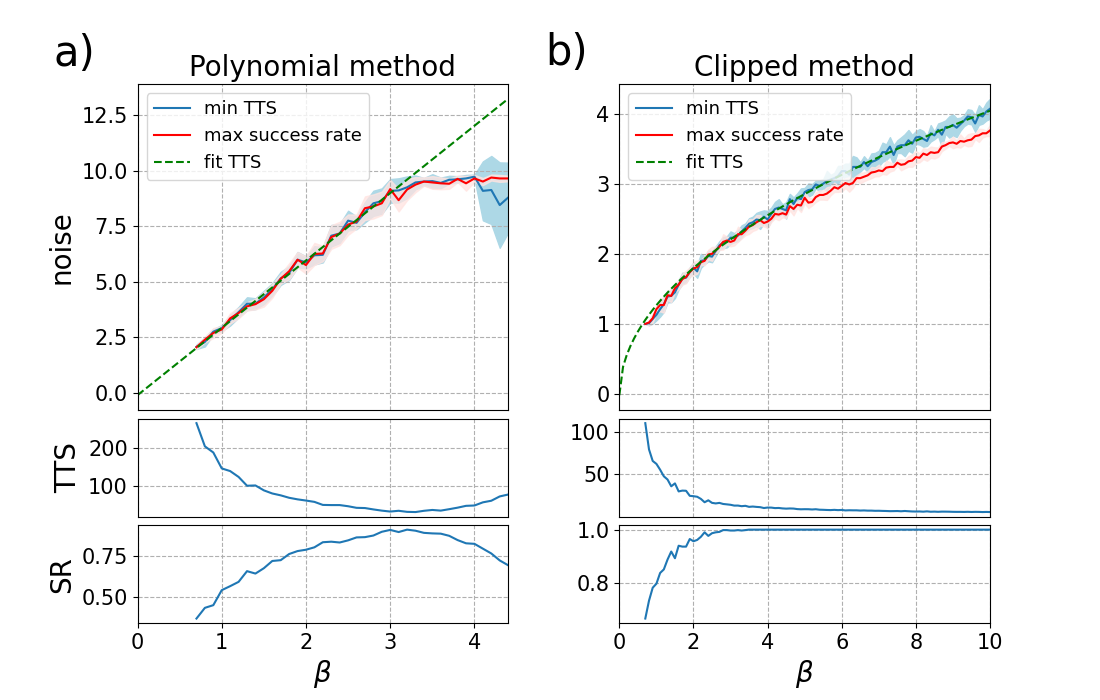}
		\caption{The noise strength which leads to the maximal success rate and the minimal TTS in function of the coupling strength $\beta$ for the g05\_60.3 problem using the polynomial \nonline (a) or clipped \nonline (b). The minimal TTS and maximal success rate (SR) are added as two small figures under the main figure. The standard deviation (shown as light colored areas) on the $\gamma$ values is determined by repeating the calculation of success rate and TTS 10 times. The relationship is fitted with a line through the origin for the polynomial \nonline and with a square root for the clipped \nonlin. Due to the scan limits, this relationship can fail for too high or low coupling strength values.}
		\label{g05_60.3_gb}
	\end{figure}

	First, we will look at small problems ($N \lesssim 100$), starting with our exemplary problem g05\_60.3 from the BiqMac library. For this problem, the relationship between $\beta$ and $\gamma$ is shown in Fig.~\ref{g05_60.3_gb}(a) for the polynomial \nonline and in Fig.~\ref{g05_60.3_gb}(b) for the clipped \nonlin. Not only the best noise value per coupling strength is given, but also the corresponding TTS and success rate (SR). To minimize randomness influencing the relationship, we have repeated the whole parameter scan 10 times. We focused on the parameter region where success was seen in Fig.~\ref{g05_60.3}, excluding too low $\beta$ values. The standard deviation on the $\gamma$ values for the minimal TTS and maximal success rate is displayed as a color spread, to give an idea of the differences between the different runs. When multiple noise strengths yield the same best result, the one with the lowest noise strength is chosen. As the noise strength is only scanned in the interval [2,10], the linear relationship flattens out when the maximum noise is reached.
	The best TTS is fitted with a straight line through the origin ($\gamma = A \beta$) for the polynomial \nonline and with a square root ($\gamma = B \sqrt(\beta)$) for the clipped \nonlin. 
	When $\beta$ increases, we see that the TTS decreases and success rate increases for both \nonlines. So the best TTS values are always found for high noise and coupling strength. It is therefore possible to optimize the TTS in Fig.~\ref{TTS_dif} even further by expanding the range to higher $\beta$ and $\gamma$, but the most spectacular decrease in TTS happens already earlier, during the transition from low ($\beta \lesssim 1$) to high coupling strengths ($\beta > 1$). The TTS and success rate starts to worsen after $\beta = 3$ for the polynomial \nonlin, but this is because the parameter scan did not include high enough noise values, as mentioned earlier.
	
	We are interested in the slope $A$ of the green-dotted fit in Fig.~\ref{g05_60.3_gb} and we investigate if it is determined by some easy determinable problem characteristics, such as $N$ or the number of connections. We assume that for the noise to make an impact on the spin amplitude, it has to have a magnitude similar to the size of the rest of the nonlinear transfer function. As both are on average zero, we look at the standard deviation.
	\begin{equation}
		\text{std}(- x_i - x_i^3 + \beta \sum_j J_{ij} x_j) \sim \text{std}(\gamma F(t) )
	\end{equation}
	We assume that $|- x_i - x_i^3| \ll |\beta \sum_j J_{ij} x_j|$, which corresponds to high $\beta$ and/or a high number of connections per spin (resulting in a large sum). Furthermore, we consider  $J_{ij} \in \{-1,0,1\}$ and enforce amplitude homogeneity ($\forall i \in \{0,1 ,... n-1\}: x = \abs{x_i} $). To simplify the notation, we will contract the sign of $J_{ij} x_j$ in  $s_j$. This results in
	\begin{equation}
		\beta  \  \text{std} (x  \sum_{j, J_{ij} \neq 0} s_j ) \sim \text{std}(\gamma F(t) )
	\end{equation} 
	When simulating problems, we see that the spins have approximately 50\% chance to be 'up' and 50\% to be 'down'. So $s_j$ is approximately binomial distributed (see section \ref{bindistr_section} in supplementary materials), so we can approximate the standard deviation as $\text{std}(\sum_{j, J_{ij} \neq 0} s_j) =  \sqrt{C}$ where $C$ is the number of connections per spin. This leads us to the simple formula
	\begin{equation} \label{theorie}
		\gamma \sim \beta \sqrt{C} x
	\end{equation}
	
	When each spin is connected to more than half the other spins ($C>N/2$) and the connections are purely anti-ferromagnetic, there are for sure connected spins which will have the same direction. More specific, each spin is connected with at least $C-N/2$ spins which have the same direction. This means that $C-N/2$ spins have a negative impact on the sum $\sum_j s_j$ and another $C-N/2$ connected spins are needed to neutralize them, making the sum $0$ again. This impacts the derived formula \eqref{theorie}, as now  $\text{std}(\sum_{j, J_{ij} \neq 0} s_j) =  \sqrt{N-C}$ and thus $\gamma \sim \beta \sqrt{N-C} x$. 
	
\begin{figure}
		\centering
		\includegraphics[width=0.75\linewidth]{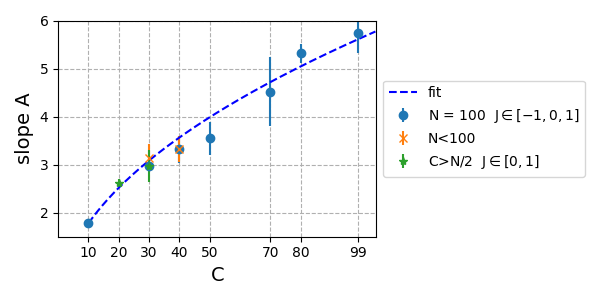}
		\caption{The slope $A$ of the linear fit on the relationship between the best noise strength $\gamma$ and coupling strength $\beta$ in function of the average number of connections per spin $C$. The problems with 100 spins are shown as blue dots, the problems with lower sizes as orange crosses and the problems with more than N/2 antiferromagnetic connections per spin as green stars. }
		\label{con_spin}
	\end{figure}

	Here, we observe that the relationship between the noise and coupling strength is determined by the number of connections per spin. As the assumptions, such as amplitude homogeneity, only hold approximately in reality, we have analyzed several MaxCut problems to confirm this relationship. We determine the slope $A$ for a range of problems with different characteristics, similar to what we did in Fig.~\ref{g05_60.3_gb}. In Fig.~\ref{con_spin}, we show the slopes $A$ corresponding to different problems in function of their number of connections per spin. The points represent the mean slope of a group of similar problems. These similar problems are simulated in the same way and have the same $C$ and $N$. Furthermore, they have a similar coupling matrix, which means that $J$ has the same coupling strengths ($J_{ij} \in \{-1,0,1\}$ or  $J_{ij} \in \{0,1\}$). The error bar represent the standard deviation on the mean slope. In Fig.~\ref{con_spin}, all the problems have 100 or less spins. We divided the problems in three groups: the problems with $N=100$ and $J_{ij} \in [-1,0,1]: \forall i,j$ (blue), the problems with $N<100$ and $J_{ij} \in [0,1]: \forall i,j$ (orange) and the problems with $N = 100$, $C>N/2$ and $J_{ij} \in [0,1]: \forall i,j$ (green). Notice that for the last group of problems, the slope $A$ is in function of $N-C$ instead of $C$, as these type of problems follow a slightly different relationship. Most problems come from the BiqMac library, but there are also self-generated problems. Like the BiqMac problems, they are generated using rdn\_graph of rudy \cite{noauthor_biq_nodate}. 
	
	The slope $A$ is not always determined in the same way. For all the problems, the relationship between $\beta$ and $\gamma$ is fitted on basis of the minimum TTS and the maximum success rate. Mostly, the fit parameters for these two were very similar. In extreme cases, such as very easy or hard problems, the TTS and success rate lead to different relationships. For easy problems, we used the minimum TTS to fit and for difficult problems, we base the fit on the maximum success rate. 
	In Fig.~\ref{con_spin}, we also show a fit of the slope $A$ in function of $C$ using Eq. \eqref{theorie}, so $A = \gamma / \beta \sim c_1 \sqrt{C} + c_2$ where $c_1$, $c_2$ are fit parameters. While the exact coupling matrix is needed to correctly deduce the optimal noise level,  we see that the amount of connections per spin predicts the ratio $\gamma / \beta$ fairly well. In Fig.~\ref{con_spin}, $c_1 = 0.56 \pm 0.03$ and $c_2 = 0.01 \pm 0.13$ with a R$^2$ of 0.97. Therefore, this scaling law can be used to predict the best noise values for a chosen $\beta$ for all random problems without local fields and with $J\in \{-1,0,1\}$. 
	
	\subsection{Large problems} \label{relatiegroot}
\begin{figure}
		\begin{subfigure}{0.49\linewidth}
			\includegraphics[width=\linewidth]{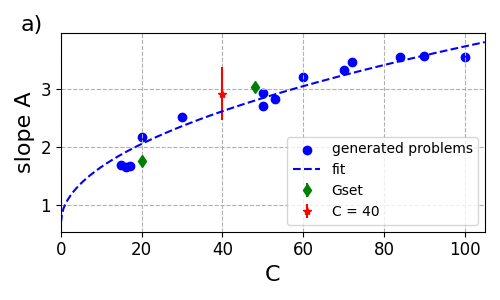}
			\caption{All large problems in function of $C$}
			\label{relatie_con_Na}
		\end{subfigure}
		\begin{subfigure}{0.49\linewidth}
			\includegraphics[width=\linewidth]{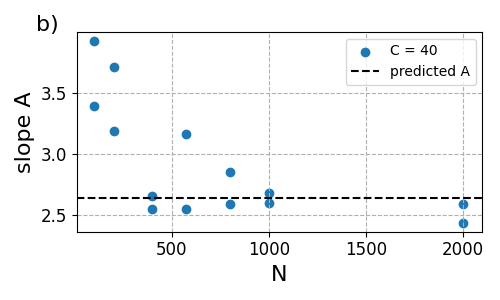}
			\caption{$C = 40$}
			\label{relatie_con_Nb}
		\end{subfigure}
		\caption{The average best $\beta$ for $\gamma \in [20,25]$ in function of $C$ (a) or $N$ (b). Similar problems are averaged and shown with a error bar, such as the Gset and our own $C = 40$ problems. The best $\beta$ is determined using the TTS (blue dots) and success rate (red diamonds) and afterwards fitted using \eqref{theorie}.\\ 
		  }
		\label{relatie_con_N}
	\end{figure}

	We now investigate MaxCut problems between 200 and 2000 spins with densities 0.5,0.1,0.06 and 0.01. We generated them using the method rnd\_graph of rudy \cite{noauthor_index_nodate,helmberg_spectral_2000,noauthor_biq_nodate}. We estimate the ground state energy relying on a Bayes Optimization of CAC IM \cite{leleu_destabilization_2019}. These problems, together with the random problems in Gset, namely G1-G10 and G22-G31, will be used to qualify the influence of the number of connections on the slope $A = \gamma/\beta$ in Fig.~\ref{relatie_con_N}(a). The similar Gset problems are grouped together. They are displayed as green diamonds and have an error bar to show the variation between the problems. The slopes $A$ are fitted using $A = \gamma / \beta = c_1 \sqrt{C} + c_2$, leading to $c_1 = 0.31 \pm 0.04$ and $c_2 = 0.7 \pm 0.3$ with a R-squared value of 0.76. 
	
	As most problems are newly generated, we cannot make use of an established best-known cut value. Additionally, our data points in Fig.~\ref{relatie_con_N}(a) results from 1 problem, due to computation time limits. In the previous fit of the slope in function of C for small problems in Fig.~\ref{con_spin}, all the data points were averages over at least 3 problems. These two issues are probably the reason why there is more variation in Fig.~\ref{relatie_con_N}(a) and the fit quality is lower than in Fig.~\ref{con_spin}. Adding more problems and determining the highest cut value more accurate would probably make this relationship more clearly visible and improve the R-squared value.

	 We also want to further verify that only the average number of connections per spin $C$ influences the relationship between the noise strength $\gamma$ and the coupling strength $\beta$. We will look at the influence of $N$ on the best $\beta$ value for a certain noise strength. In Fig.~\ref{relatie_con_N}(b), self-generated problems with $C = 40$ but different problem sizes are shown in function of $N$. As reference, we also displayed their predicted slope $A$ calculated using the fit in Fig.~\ref{relatie_con_N}(a) as a black dotted line. The size of the problem does not seem to influence the best parameter value, but nonetheless the variation on the best coupling strength $\beta$ is quite large in comparison to the Gset problems, as can be seen in Fig.~\ref{relatie_con_N}(a). We speculate that this difference in variation is due to how well the ground state energy is known: as the Gset problems are often used, the best-known cut value probably approaches the optimum cut value quite well. The highest cut value of the self-generated problems was only estimated roughly by us using CAC. We often encountered higher cut values during the simulations with all models. Moreover, for smaller problem sizes, the variation seems to increase, probably because they are more prone to statistic variations. 
	
	\subsection{Predicting the best parameter set}\label{verify}
\begin{figure}
		\centering
		\includegraphics[width=\linewidth]{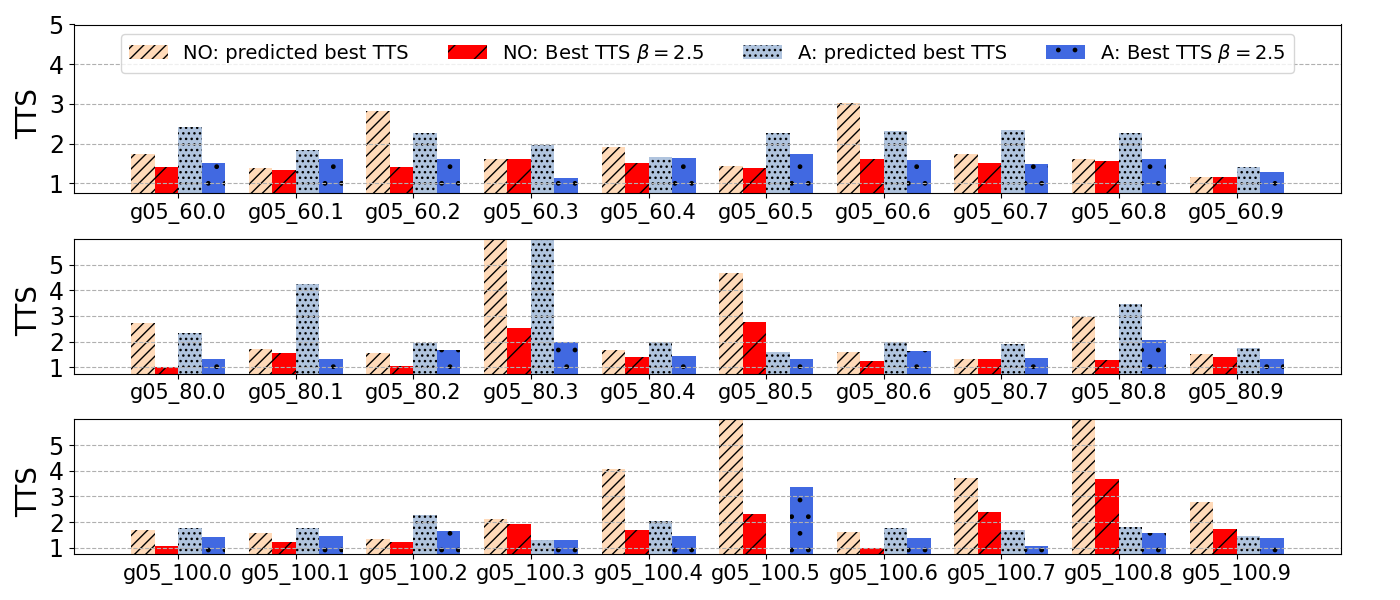}
		\caption{The TTS using the predicted noise value for $\beta = 2.5$  and the best TTS for $\beta = 2.5$ relative to the best TTS using the same \nonline for BiqMac problems. The two left red striped bars are for the \noisee (abbreviated as NO in the legend) and the two right blue dotted bars represent annealing (abbreviated as A in the legend). For both annealing and noise, the TTS is relative to the best TTS for that problem and \nonlin. As we want the predicted TTS to be as good as possible, we want our relative TTS to be close to 1. }
		\label{TTS_predicta}
	\end{figure}

	Both the small and the large problems fit nicely with the theoretical relationship \eqref{theorie}. Here, we want to analyze if we are able to predict the ideal noise value using the two fits from Fig.~\ref{con_spin} and Fig.~\ref{relatie_con_N}(a)). First, we will try to predict the noise leading to a low TTS for the small problems where the success rate is based on the optimal cut value. We choose $\beta = 2.5$ and then predict the best noise value using the fitted relationship from Fig.~\ref{con_spin}. We choose $\beta = 2.5$ as this is already large enough to improve the TTS drastically and is not too big to lead to numerical instabilities in the simulations. Nonetheless, choosing $\beta$ a bit higher or lower would provide similar results. In Fig.~\ref{TTS_predicta}, we show the TTS for the predicted noise value relative to the best TTS for each g05 BiqMac problem. To see the influence of our choice for the coupling strength $\beta$, we compare this TTS, which used the fit from Fig.~\ref{con_spin}, to the best TTS that can be found for $\beta = 2.5$. Using the predicted noise value results in TTS values which are on average the double of the best TTS. The difference with the best TTS for $\beta = 2.5$ is even smaller, indicating that our prediction of the best noise for a certain $\beta$ value works fine. Nonetheless, there are some problems where the best TTS is drastically lower than the TTS for the chosen $\beta$ and predicted $\gamma$, such as for problem g05\_100.5, where there even is no success when using the predicted noise value for annealing. Due to the overall low success rate for this problem, the TTS does not change smoothly anymore, making it harder to predict. The same occurs for the problem g05\_80.3. For problem g05\_100.8, the predicted TTS is only double of the best TTS for $\beta = 2.5$, but the TTS can be much improved by choosing a higher $\beta$ value.
	
\begin{figure}
		\centering
		\includegraphics[width=\linewidth]{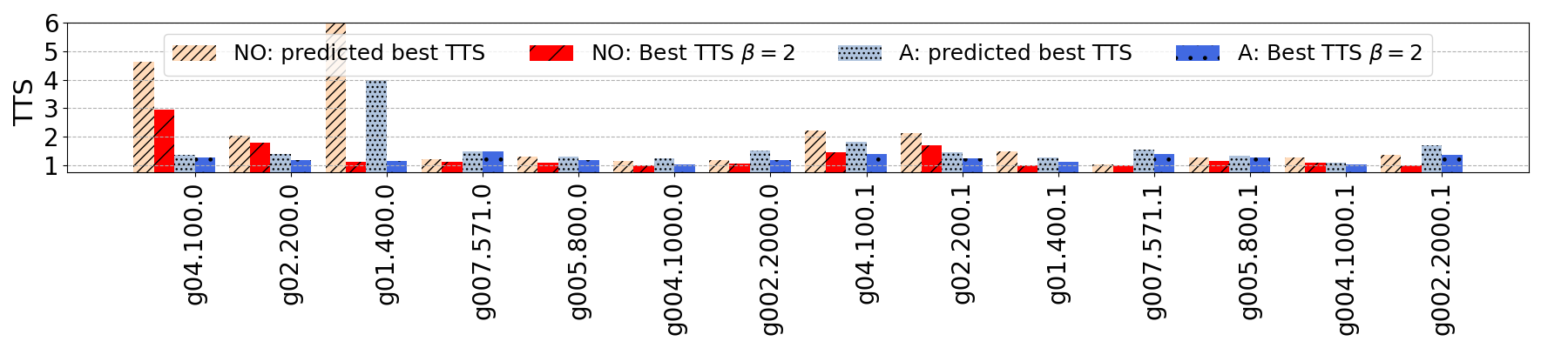}
		\caption{The TTS using the predicted noise value for $\beta = 2$  and the best TTS for $\beta = 2$ relative to the best TTS using the same \nonline for the $C = 40$ problems. The two left red striped bars are for the \noisee and the two right blue dotted bars represent annealing. For both annealing and noise, the TTS is relative to the best TTS for that problem and \nonlin. As we want the predicted TTS to be as good as possible, we want our relative TTS to be close to 1. }
		\label{TTS_predictb}
	\end{figure}

	The same comparison is done in Fig.~\ref{TTS_predictb} for the $C = 40$ problem from Fig.~\ref{relatie_con_N}(b) to verify the fit for the large problems. Here we take $\beta$ fixed on $2$ and use the fit in Fig.~\ref{relatie_con_N}(a) to predict $\gamma$. We choose a different $\beta$ value, as the predicted noise values for higher $\beta$ values fell outside of the scanned noise range. Similar to the BiqMac problems, we see that using the fitted relationship allows us to find a decent parameter set which gives TTS values that are very close to the best TTS. For annealing, the TTS is even never larger than twice the best TTS. So while there was a wide variation on the slope $A$ in Fig.~\ref{relatie_con_Nb}, predicting the best parameter set using the slope of the fit leads to good results. 

	\section{Discussion and conclusion} \label{discussion}
	We have investigated the importance of injecting noise in an IM when it evolves to the ground system. It is clear that choosing the right noise strength and coupling strength can improve the TTS and success rate immensely. By comparing the annealing scheme and the \noisee to CAC, we observed that adding large noise could make these models competitive. 
	While the \noisee performs faster for smaller problems, annealing with large noise is the way to go for larger problems. 
	
	Selecting the noise strength can  be done on basis of the connectivity, eliminating the need of scanning the noise. The estimation of the noise strength uses the fits of the theoretical correlation $\gamma \sim \sqrt{C} \beta$ and is verified numerically. Using the fit parameters found in Fig.~\ref{con_spin} and in Fig.~\ref{relatie_con_N}(a), it is possible to predict a good noise value $\gamma$ for a chosen coupling strength $\beta$ for all problems similar to the ones used for the fit. So for new random problems with $J \in \{-1,0,1\}$ and without local field, it is possible to get a good TTS using the found fits. While the exact ideal noise strength $\gamma$ depends on the exact coupling matrix, estimating it roughly using the number of connections per spin leads to TTS which are around the double of the best TTS, as can be seen for the Biq Mac MaxCut in Fig \ref{TTS_predicta} and the self-generated $C = 40$ problems in Fig \ref{TTS_predictb}.   
	
	We always focused on the relationship between the noise strength $\gamma$ and coupling strength $\beta$, ignoring the annealing cases. Nonetheless, annealing is faster and has a higher success rate for more difficult problems. We choose to focus on the relationship between $\gamma$ and $\beta$ as the relation between the annealing speed $v$ and the noise strength $\gamma$ can be derived from it: $v = \beta / t_A$ where $t_A$ is the average time needed to reach the ground state. For the verification of the fits of the slope with the number of connections per spin in Fig \ref{TTS_predicta} and Fig.~\ref{TTS_predictb}, we use that $t_A = 50$, but any value between 25 - 75 can work. Furthermore, there is always a larger spread on the good parameter values for annealing, as can already be seen in Fig \ref{g05_60.3}. This phenomena is even more prominent for larger problems, such as the Gset, where annealing has lower TTS and the success rate is overall lower than for BiqMac problems. 
	
	In this study, the focus was on the polynomial \nonline and the clipped \nonlin, even when it is already shown that other \nonlins, in particular the periodic \nonlin, work better. We limited ourselves to \nonlines which could be explored using only a single parameter, but we suspect that injecting noise is also beneficial for the other \nonlins. Especially, we expect the behaviour of the periodic \nonline to be similar to the clipped \nonlin.
	Furthermore, it could also be interesting to look at other types of coupling matrices, with different coupling strengths. While a combination of ferro- and antiferromagnetic relationships does not seem to influence the required noise much, we suspect it will be not the case when using different coupling strengths. We conjecture that an extra term containing the standard deviation of $J$ needs to be included in the relationship in Eq. \ref{theorie}. Once $J$ is correctly included in the relation, the found fits can then be used to predict the best noise value for various types of problems, without the need for further fitting or scanning the noise strength. This would make it possible to solve new problems efficiently without the need for extensive parameter scans beforehand. 
	
	\section{Methods}
	All the results come from numerical simulations using gradient descent with a maximum of 10000 - 20000 iterations and a time step of $dt = 0.005$. For all the simulations, we made use of the transient success rate, repeating the simulations 100 times for the g05 problems, 200 for our own problems and 500 times for the Gset. The Gset simulations ran always for 15000 iterations to find the lowest energy or highest cut value. Noise is injected every time step as explained in \cite{san_miguel_stochastic_2000} and is also clipped for the clipping nonlinearity. The best CAC TTS is determined using BayesOpt \cite{martinez-cantin_bayesopt_2014} with 40 initial samples and 120 iterations. For the Gset, $\beta$ is taken as a parameter in the optimization. For the BiqMac problems, an additional sweep over $\beta$ is performed for 50 values between 0 and 5. The other best TTS are determined by a parameter scan of 50x50 (BiqMac) or 25x25 (Gset and self-generated problems) over noise and coupling strength. 
	
	The time-to-solution is calculated as following
	\begin{equation}
		TTS = \frac{1}{N_{success}} \sum_i t_i \frac{\log{0.01}}{\log{1-success rate}}
	\end{equation}
	where $N_{success}$ is the number of times the ground state energy was reached and $t_i$ the simulation time (number of iterations $\times$ time step) it took in the i$^{th}$ successful run to reach the ground state energy. In the literature, other definitions for TTS are also used, such as 
	\begin{equation}
		TTS =  min_{t_R} t_R \frac{\sqrt{0.01}}{\sqrt{1-success rate}}
	\end{equation}
	As this makes the calculations of the TTS more cumbersome and does not provide any additional insight, this definition will not be used. In general, this definition leads to TTS which are the double of the first definition. 
	
	The problems are MaxCut problems from the BiqMac library \cite{noauthor_biq_nodate} and from the Gset library \cite{noauthor_gset_nodate}. More information about the Gset is given in \cite{matsuda_benchmarking_2019}, from where we got the best known solutions. 
	
	\section{Data availability}
	The authors declare that all relevant data are included in the manuscript. Additional data are available from the corresponding author upon reasonable
	request.
	\section{Author contributions}
	L.M. performed the simulations and wrote the manuscript. G.V. and G.V.d.S. supervised the	project. All authors discussed the results and reviewed the manuscript.
	\section{Additional information}
	\textbf{Competing interests:} The authors declare no competing interests.\\
	\textbf{Acknowledgements:} 
	The computational resources and services used in this work were provided by the VSC (Flemish Supercomputer Center), funded by the Research Foundation Flanders (FWO) and the Flemish Government – department WEWIS.

	\section{Supplementary materials}
	\subsection{Clipped \nonline} \label{clip}
\begin{figure}
		\includegraphics[width=\linewidth]{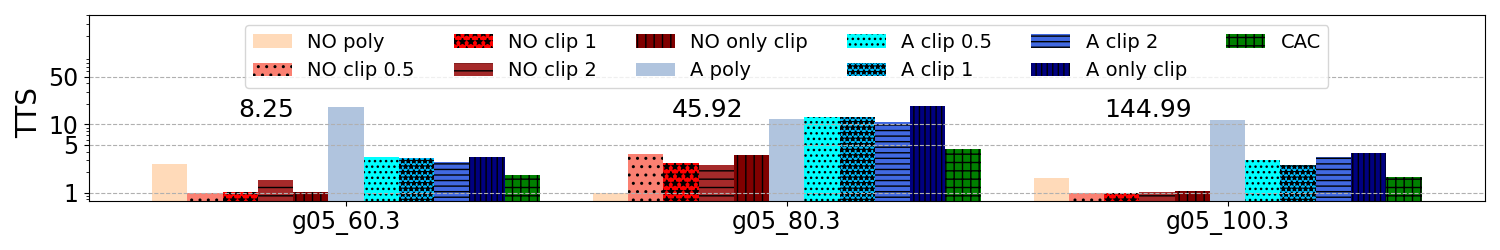}
		\caption{The TTS of the \noisee (red, stripes) and annealing (blue) scheme for the polynomial \nonline and 4 different clipped \nonlines ($x_{cutoff} \in \{ 0.5,1,2\}$ and $x_{cutoff}=1$ when $-x_i^3$ is omitted) relative to the best TTS. The relative TTS for CAC is shown as reference. In the legend, the several \nonlines are abbreviated: NO stands for the \noisee, A for the annealing scheme, poly for the polynomial \nonline and clip c for the clipped \nonline with $x_{cutoff} = c$. }
		\label{TTS_alle_methods}
	\end{figure}

\begin{figure}[H]
		\centering
		\includegraphics[width=\linewidth]{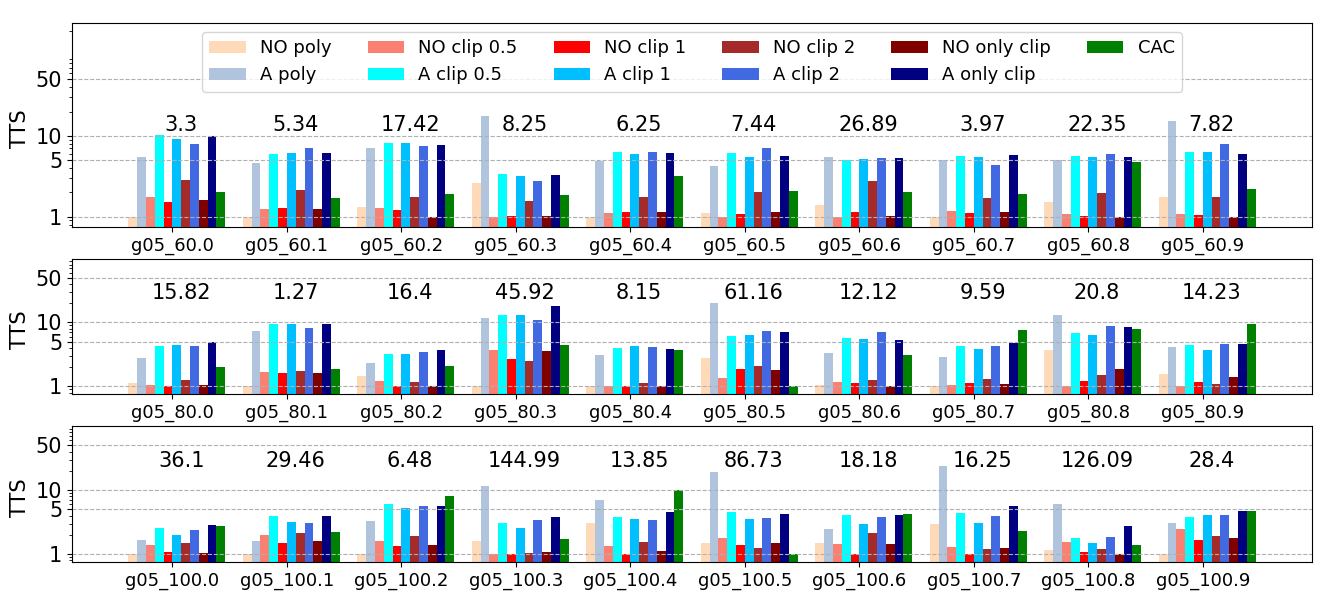}
		\caption{The TTS of the \noisee (red) and annealing scheme (blue) of the polynomial \nonlin, several clipped \nonlines ($x_{cutoff} \in \{ 0.5,1,2\}$ with $-x_i^3$ included and $x_{cutoff}=1$ without $-x_i^3$) and CAC  relative to the best TTS of these models. The TTS of the best model is mentioned above to give an idea of the timescale.  In the legend, the several \nonlines are abbreviated: NO stands for the \noisee, A for the annealing scheme, poly for the polynomial \nonline and clip c for the clipped \nonline with $x_{cutoff} = c$.}
		\label{TTS_dif_extra}
	\end{figure}

	We compared the TTS for several clipped \nonlins: we investigate the influence of $x_{cutoff}$ by taking $x_{cutoff} \in \{0.5,1,2\}$. Additionally, we looked at the influence of the $-x^3_i$ in equation \eqref{clipped} for $x_{cutoff} = 1$. This is shown in Fig.~\ref{TTS_alle_methods} for 3 exemplary BiqMac problems of 60 (g05\_60.3), 80 (g05\_80.3) and 100 (g05\_100.3) spins. As a reference, the results for the polynomial \nonline and CAC (green) is also added. The whole comparison for the TTS of all g05 BiqMac problems can be found in Fig.~\ref{TTS_dif_extra}. Like Fig.~\ref{TTS_alle_methods}, it shows the TTS for 11 different models and both figures have the same color code for the models. While in Fig.~\ref{TTS_alle_methods}, the models are grouped per scheme, in Fig.~\ref{TTS_dif_extra}, the models are grouped per \nonlin. 
	
	Both figures deliver the same message: Fig.~\ref{TTS_dif_extra} shows all the compared problems, while Fig.~\ref{TTS_alle_methods} is added for better readability. 
	While the clip values and the different clipped \nonlines influence the TTS, the differences in TTS are modest and mostly fall within the expected variability due to the limited number of repetitions. Additionally, using excessively high clip values is generally counterproductive; the clip value 2 almost always leads to a higher TTS. Therefore, we only looked at $x_{cutoff} = 1$ with the term $-x_i^3$ included in the differential equation for all the problems.
	Furthermore, the place of the parameter set that leads to the best TTS differs depending on $x_{cutoff}$. How higher $x_{cutoff}$, how higher the noise has to be for the same coupling strength. We did not look further into the influence of $x_{cutoff}$ on the relation, as changing $x_{cutoff}$ does not change the TTS substantially. Therefore, it will be enough to check the relationship and the influence of $C$ for one value of $x_{cutoff}$.
	
	\subsection{Influence of noise for extra problems} \label{extraprob}
	In section \ref{influencenoise}, we looked at problem g05\_60.3 to derive some results about the best system parameters to give a small TTS and to find a relationship between $\gamma$ and $\beta$ (or $v$). Here we give the results for 9 extra problems of the BiqMac library with different problem size ($N \in \{60,80,100\}$). The figures \ref{g05_N.0}, \ref{g05_N.1} and \ref{g05_N.2} are build up analog to Fig.~\ref{g05_60.3}. The TTS and success rate are shown for the \noisee and the annealing scheme for both the polynomial and clipped \nonlin. For these 9 problems, we also see a linear relationship between $\gamma$ and $\beta$ or $v$ for the polynomial \nonline and a square root dependence between $\gamma$ and $\beta$ or $v$ for the clipped \nonlin.
	\begin{figure}[H]
		\begin{subfigure}{0.99\linewidth}
			\includegraphics[width=\linewidth]{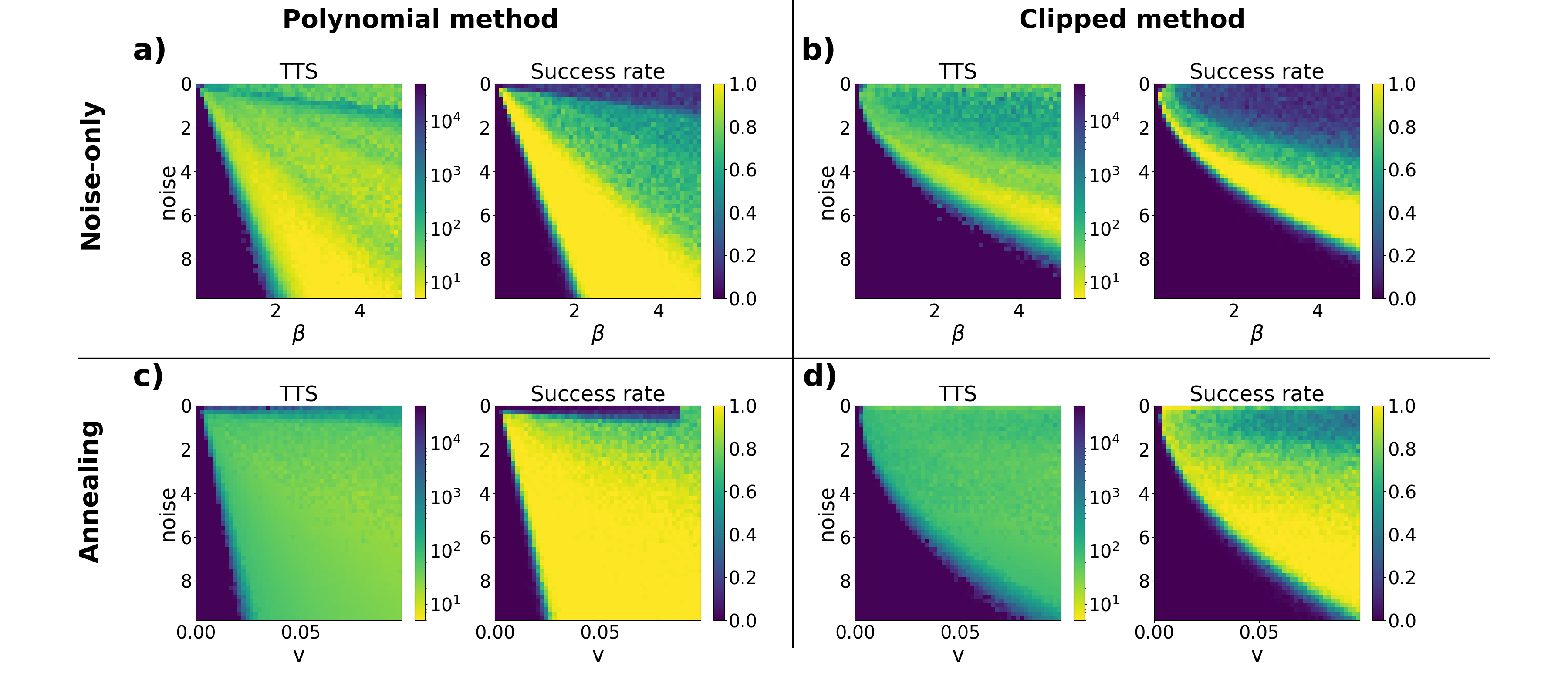}
			\caption{g05\_60.0}
		\end{subfigure}
		\begin{subfigure}{0.99\linewidth}
			\includegraphics[width=\linewidth]{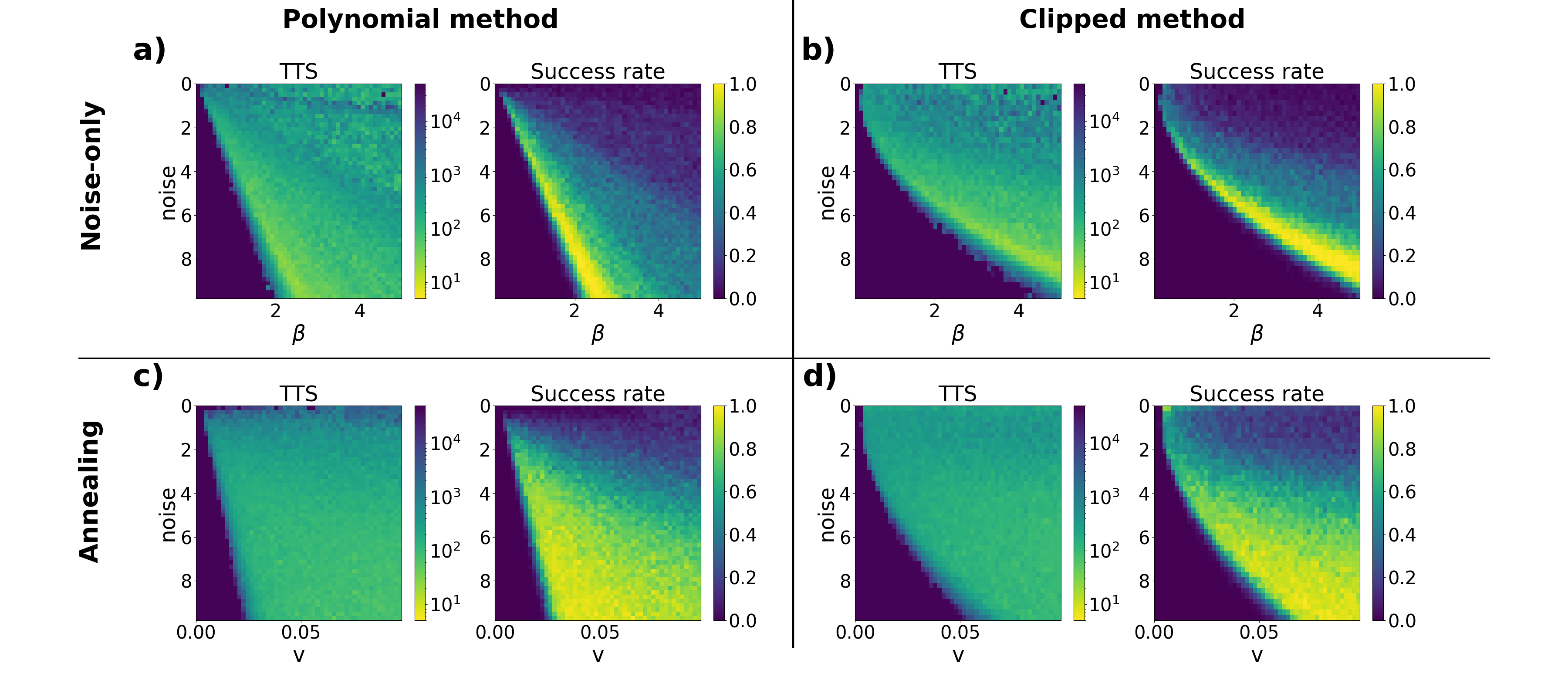}
			\caption{g05\_80.0}
		\end{subfigure}
		\begin{subfigure}{0.99\linewidth}
			\includegraphics[width=\linewidth]{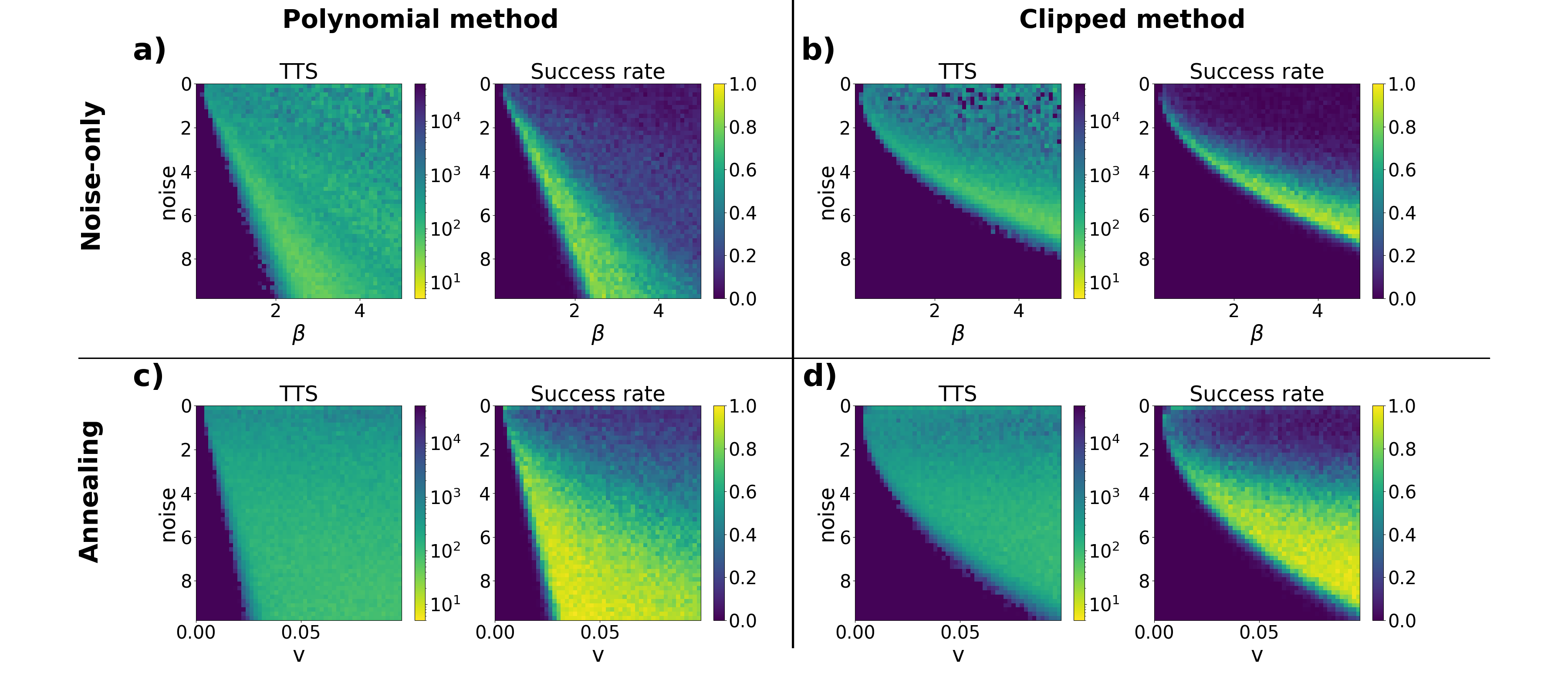}
			\caption{g05\_100.0}
		\end{subfigure}
		\caption{he success rate and the time-to-solution in function of $\gamma$  and $\beta$ for the \noisee (a-b) and $\gamma$ and $v$ for annealing (c-d) as heat map for 3 problems. From top to bottom, these problems are g05\_60.0, g05\_80.0 and g05\_100.0 from the BiqMac library. The results for the polynomial \nonline can be seen at the left (a,c) and the results for the clipped \nonline at the right (b,d). }
		\label{g05_N.0}
	\end{figure}
	
	\begin{figure}[H]\ContinuedFloat
		\begin{subfigure}{0.99\linewidth}
			\includegraphics[width=\linewidth]{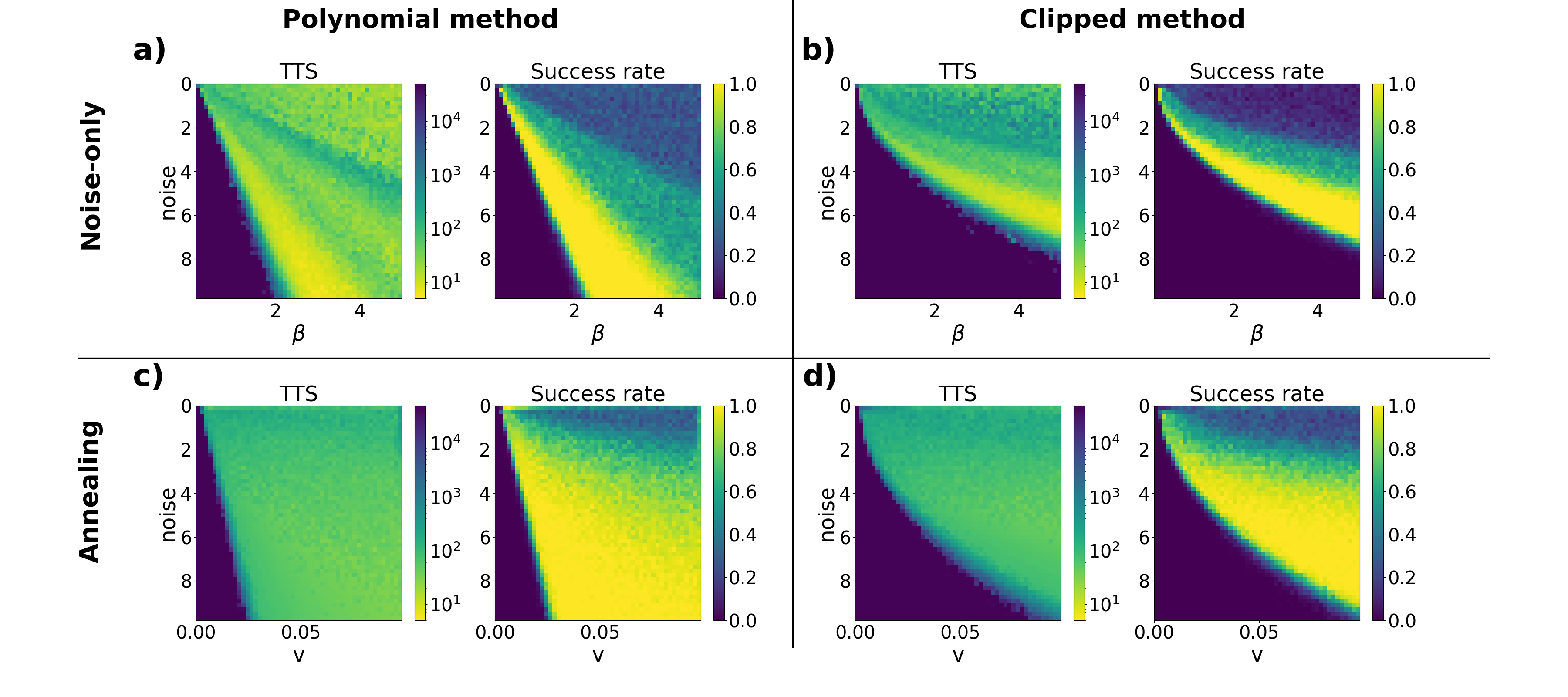}
			\caption{g05\_60.1}
		\end{subfigure}
		\medskip
		\begin{subfigure}{0.99\linewidth}
			\includegraphics[width=\linewidth]{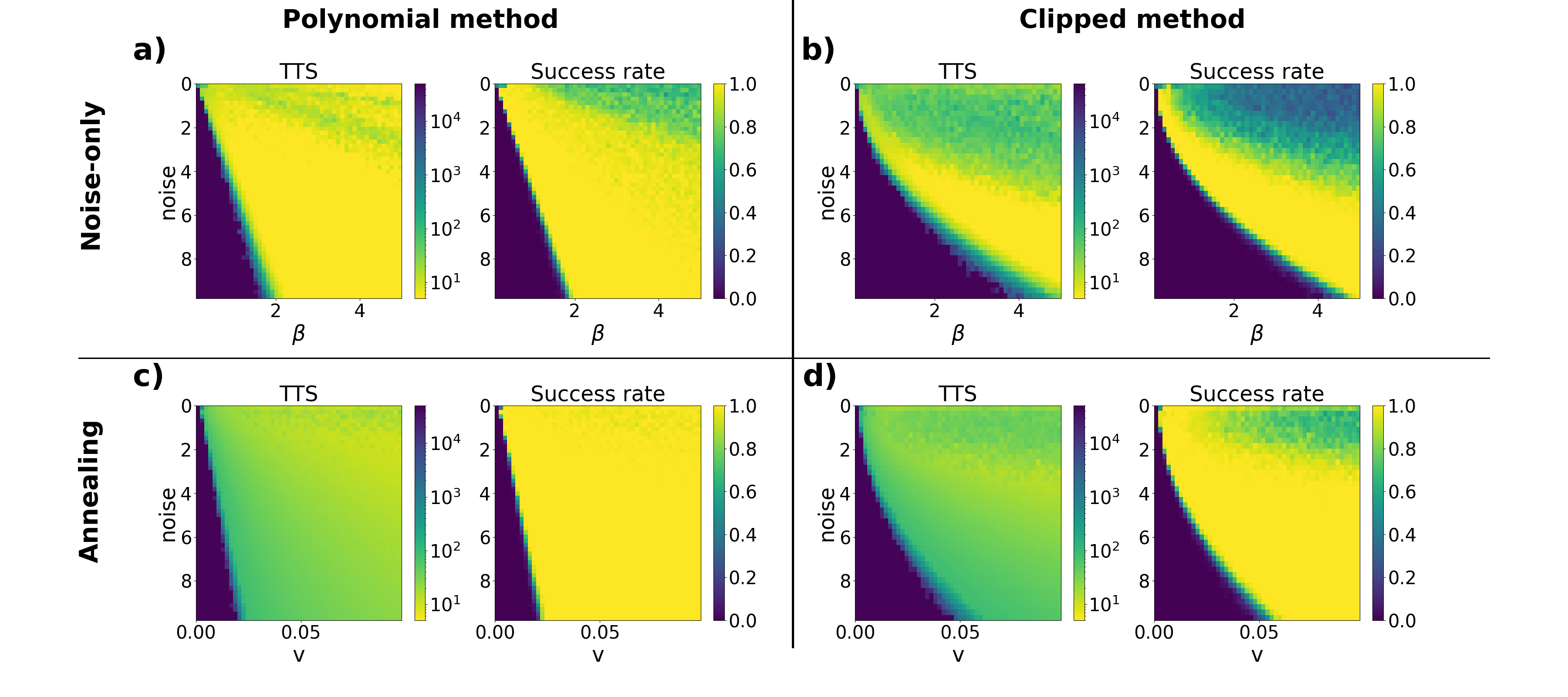}
			\caption{g05\_80.1}
		\end{subfigure}
		\medskip
		\begin{subfigure}{0.99\linewidth}
			\includegraphics[width=\linewidth]{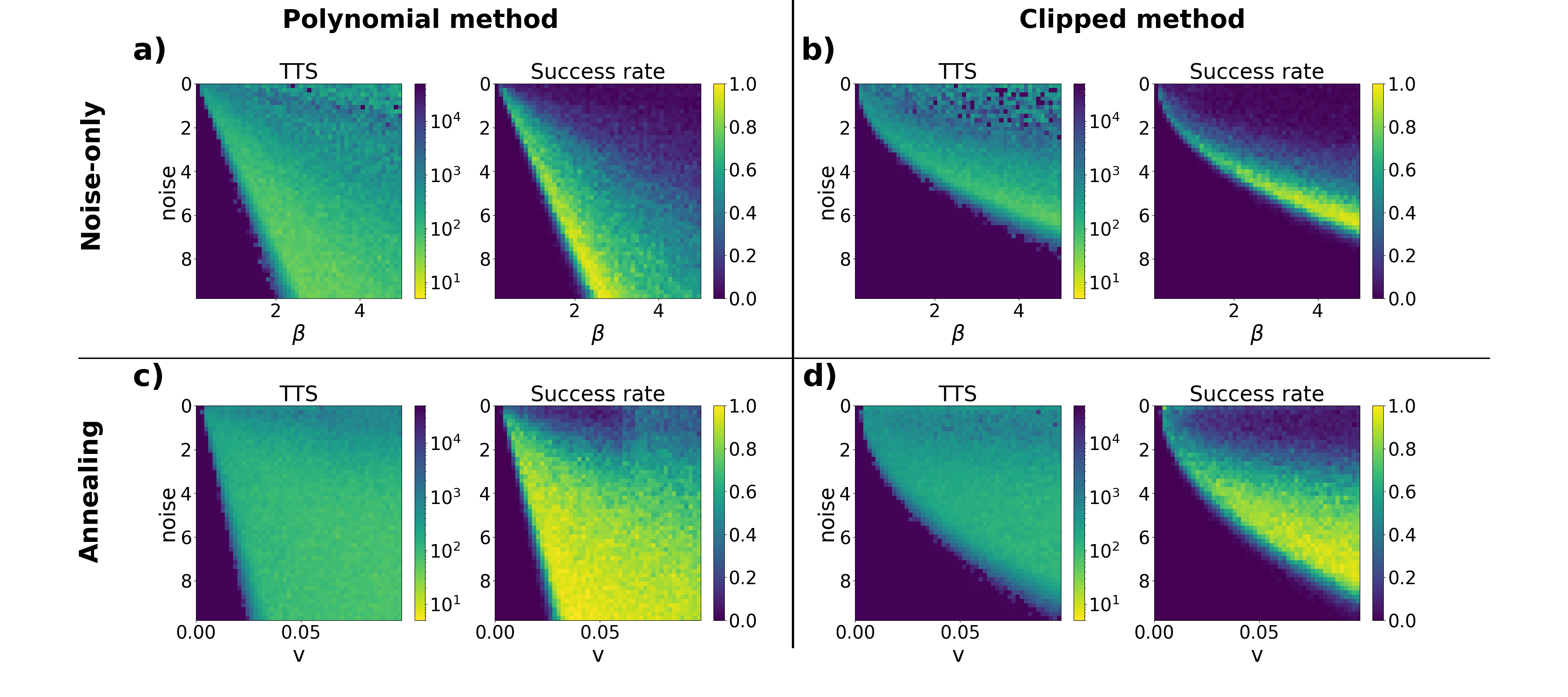}
			\caption{g05\_100.1}
		\end{subfigure}
		\caption{he success rate and the time-to-solution in function of $\gamma$  and $\beta$ for the \noisee (a-b) and $\gamma$ and $v$ for annealing (c-d) as for 3 problems. From top to bottom, these problems are g05\_60.1, g05\_80.0-1 and g05\_100.0-1 from the BiqMac library. The results for the polynomial \nonline can be seen at the left (a,c) and the results for the clipped \nonline at the right (b,d).}
		\label{g05_N.1}
	\end{figure}
	
	\begin{figure}[H]\ContinuedFloat
		\begin{subfigure}{0.99\linewidth}
			\includegraphics[width=\linewidth]{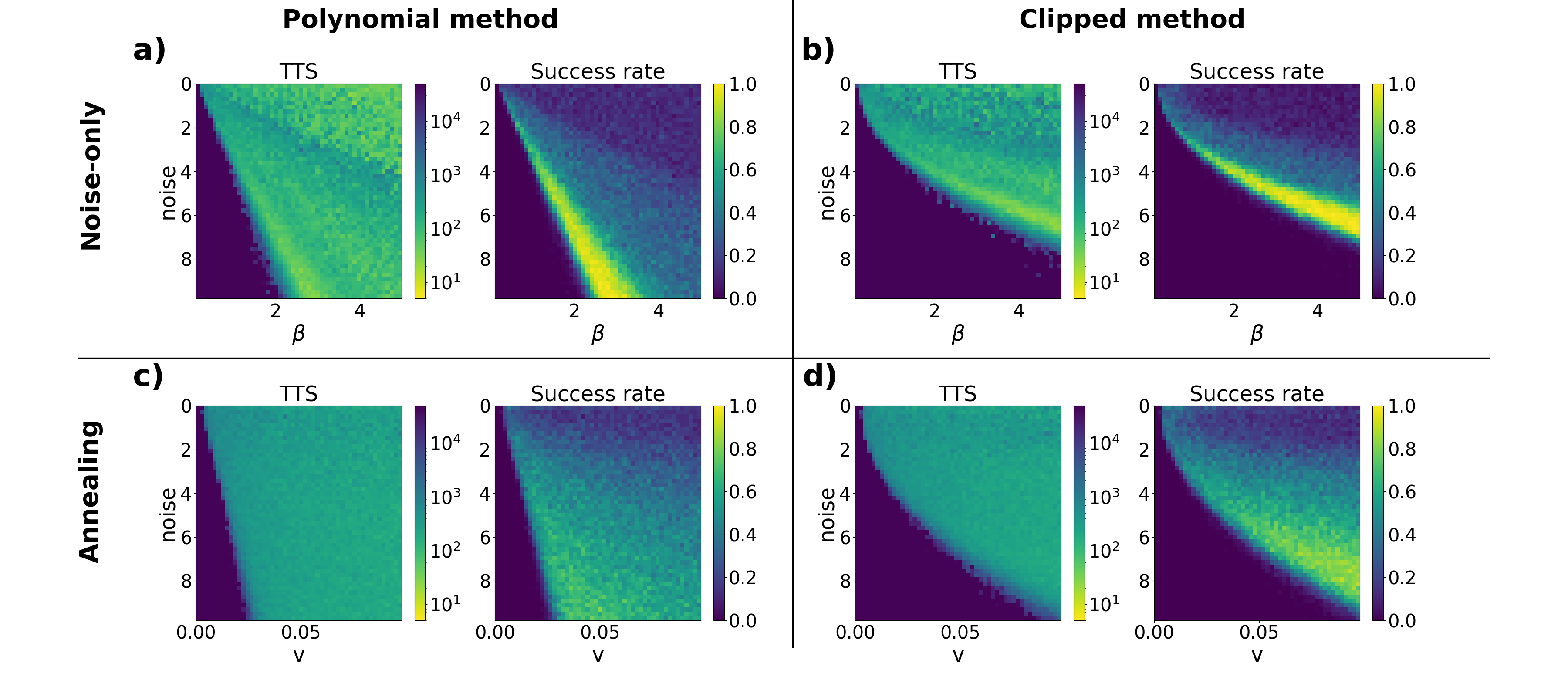}
			\caption{g05\_60.2}
		\end{subfigure}
		\medskip
		\begin{subfigure}{0.99\linewidth}
			\includegraphics[width=\linewidth]{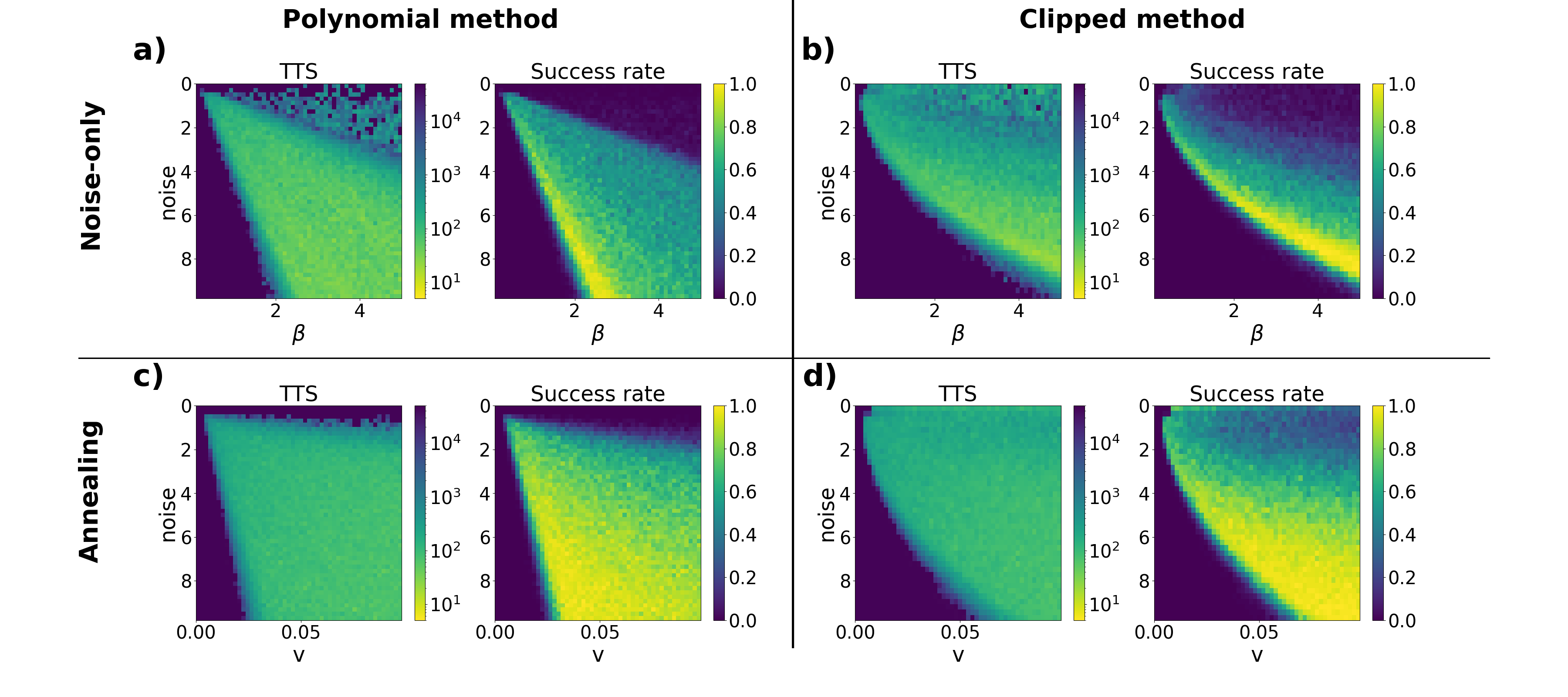}
			\caption{g05\_80.2}
		\end{subfigure}
		\medskip
		\begin{subfigure}{0.99\linewidth}
			\includegraphics[width=\linewidth]{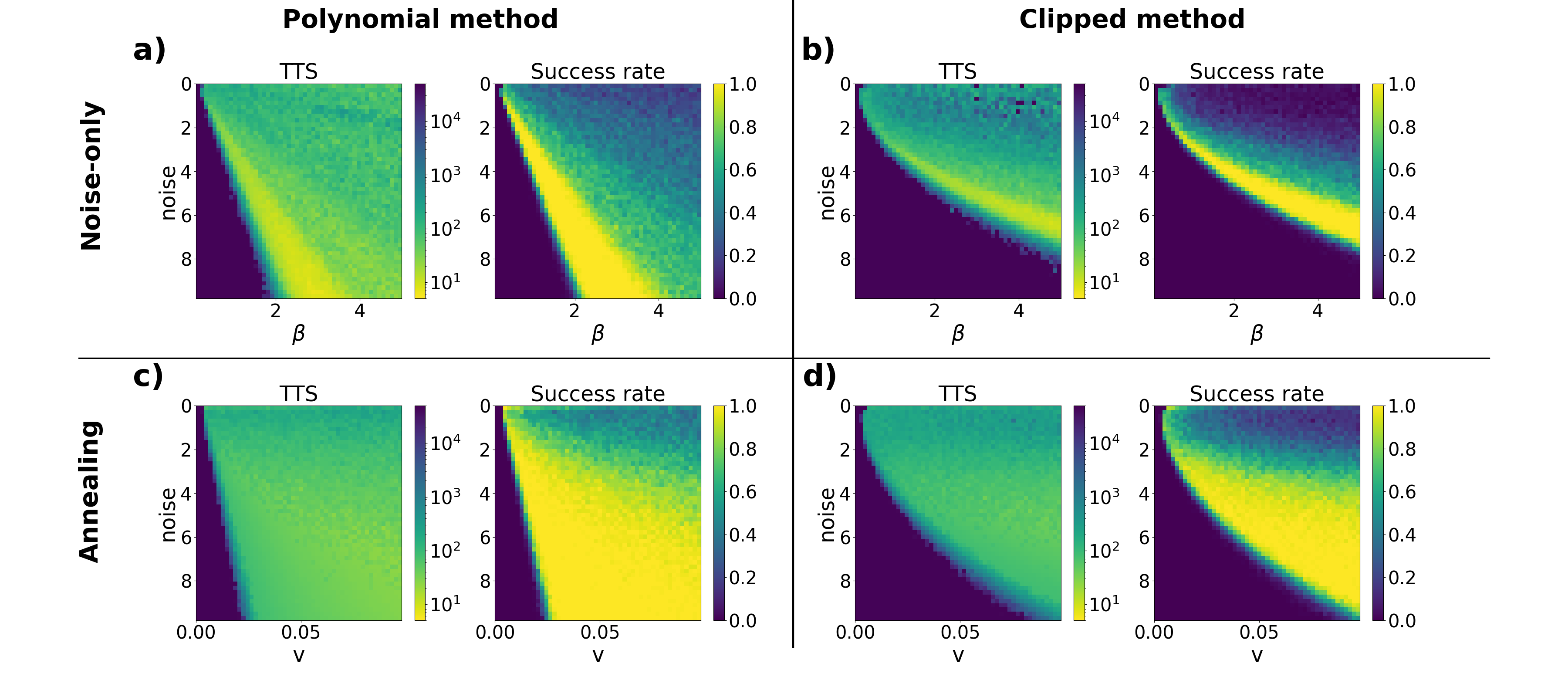}
			\caption{g05\_100.2}
		\end{subfigure}	
		\caption{he success rate and the time-to-solution in function of $\gamma$  and $\beta$ for the \noisee (a-b) and $\gamma$ and $v$ for annealing (c-d) as heat map for 3 problems. From top to bottom, these problems are g05\_60.2, g05\_80.2 and g05\_100.2 from the BiqMac library.  The results for the polynomial \nonline can be seen at the left (a,c) and the results for the clipped \nonline at the right (b,d).}
		\label{g05_N.2}
	\end{figure}
	
	 \subsection{The binomial distribution} \label{bindistr_section}
\begin{figure}
	 	\includegraphics[width=\linewidth]{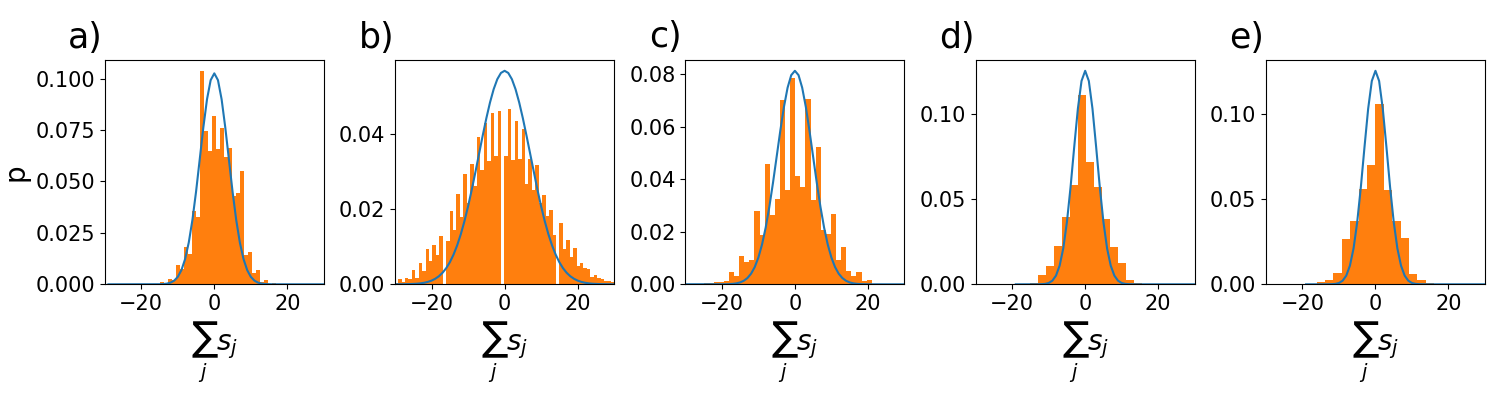}
	 	\caption{The distribution of $\sum_j s_j$ for 5 different problems: a) BiqMac problem g05\_60.3 b) BiqMac problem pm1d\_100.0 c) Gset problem G1 d) Gset problem G22 e) Gset problem G27. The blue line gives the expected binomial distribution for these respective problems.}
	 	\label{bindistr}
	 \end{figure}

	 During the derivation of the relationship between $\beta$ and $\gamma$ in section \ref{relatiesmall}, we reason that $\sum_j s_j$ follows approximately a binomial distribution. We conclude this from the fact that the spin signs have equal chance to be 1 and -1. In an binomial distribution, $s_j$ have to be independent events, but here they are connected via the coupling matrix. However, the added noise will make them more independent and we are looking over all local minima, so then we will be able to approximate them as independent events. To verify this claim, we check it for 5 different MaxCut problems using Boltzmann sampling for  $\sum_j s_j$. The resulting distribution and the expected binomial distribution are shown in Fig.~\ref{bindistr}.

	\bibliographystyle{ieeetr} 
	\bibliography{Noise_lib}
	
\end{document}